\documentclass[preprint,12pt]{elsarticle}

\usepackage{amssymb}
\usepackage{graphicx} 
\graphicspath{{}} 
\usepackage{subcaption}
\usepackage{algorithm}
\usepackage{algorithmic}
\usepackage{nomencl}  
\usepackage{amsmath} 
\makenomenclature

\journal{Sustainable Energy, Grids and Networks}

\begin{document}

\begin{frontmatter}



\title{Effect of electric vehicles, heat pumps, and solar panels on low-voltage feeders: Evidence from smart meter profiles
}


\author[inst1,inst2]{Thijs Becker\corref{cor1}}
\cortext[cor1]{Corresponding author}
\ead{thijs.becker@vito.be}
\author[inst3]{Raf Smet}
\author[inst3]{Bruno Macharis}
\author[inst1,inst2]{Koen Vanthournout}

\affiliation[inst1]{
organization={Flemish Institute for Technological Research (VITO)},
addressline={Boeretang 200}, 
city={Mol},
postcode={2400}, 
country={Belgium}
}

\affiliation[inst2]{
organization={EnergyVille},
addressline={Thor Park 8130}, 
city={Genk},
postcode={3600}, 
country={Belgium}
}

\affiliation[inst3]{
organization={Fluvius},
addressline={Brusselsesteenweg 199}, 
city={Melle},
postcode={9090}, 
country={Belgium}
}

\begin{abstract}

Electric Vehicles (EVs), Heat Pumps (HPs) and solar panels are Low-Carbon Technologies (LCTs) that are being connected to the Low-Voltage Grid (LVG) at a rapid pace. One of the main hurdles to understand their impact on the LVG is the lack of recent, large electricity consumption datasets, measured in real-world conditions. We investigated the contribution of LCTs to the size and timing of peaks on LV feeders by using a large dataset of 42,089 smart meter profiles of residential LVG customers. These profiles were measured in 2022 by Fluvius, the Distribution System Operator (DSO) of Flanders, Belgium.  The dataset contains customers that proactively requested higher-resolution smart metering data, and hence is biased towards energy-interested people. LV feeders of different sizes were statistically modeled with a profile sampling approach. For feeders with 40 connections, we found a contribution to the feeder peak of 1.2 kW for a HP, 1.4 kW for an EV and 2.0 kW for an EV charging faster than 6.5 kW. A visual analysis of the feeder-level loads shows that the classical duck curve is replaced by a night-camel curve for feeders with only HPs and a night-dromedary curve for feeders with only EVs charging faster than 6.5 kW.  Consumption patterns will continue to change as the energy transition is carried out, because of e.g. dynamic electricity tariffs or increased battery capacities. Our introduced methods are simple to implement, making it a useful tool for DSOs that have access to smart meter data to monitor changing consumption patterns.
\end{abstract}

\begin{graphicalabstract}
\includegraphics[width=1.\textwidth]{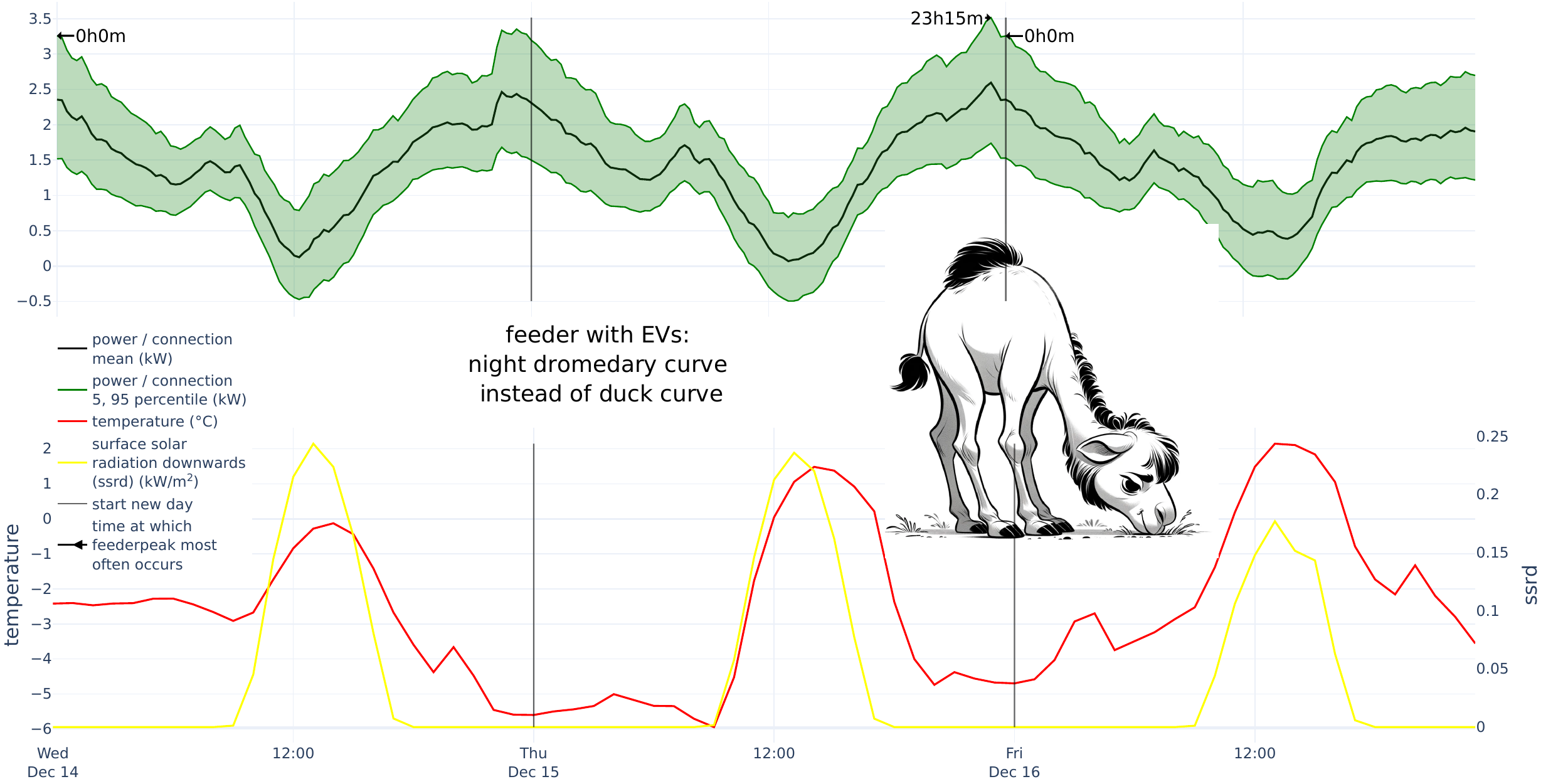}
\end{graphicalabstract}

\begin{highlights}
\item 42,089 smart meter records from 2022, 862 with heat pump, 1924 with EV, 76\% with PV.
\item Simple yet insightful methods to understand feeder-level peak loads.
\item Estimation of contribution to feeder peak of heat pump (1.2 kW) and EV (1.4 kW).
\item Duck curve becomes night-camel curve (heat pumps) or night-dromedary curve (EVs).
\item Offtake and injection peaks on feeders are of similar magnitude.
\end{highlights}


\end{frontmatter}


\nomenclature{ADMD}{After Diversity Maximum Demand}
\nomenclature{DSO}{Distribution System Operator}
\nomenclature{EV}{Electric Vehicle}
\nomenclature{HP}{Heat Pump}
\nomenclature{LCT}{Low-Carbon Technology}
\nomenclature{LVG}{Low-Voltage Grid}
\nomenclature{PV}{Photovoltaic}
\nomenclature{ssrd}{Surface Solar Radiation Downwards}
\printnomenclature

\section{Introduction}\label{sec:sample1}

Relying on fossil fuels to power our society has severe downsides, such as climate change \cite{IPCC2023} and air pollution \cite{VOHRA2021110754}. A timely transition away from fossil fuels requires the electrification of large parts of the energy system \cite{iea2023outlook}. At the residential level, which significantly contributes to greenhouse gas emissions \cite{eurostat2020}, this implies the large-scale and fast-paced installation of solar panels (PV) to generate electricity, Heat Pumps (HPs) for heating buildings and domestic hot water, and charging points for Electric Vehicles (EVs) that replace internal combustion engine vehicles.
These Low-Carbon Technologies (LCTs) have a significant effect on the Low-Voltage Grid (LVG) \cite{DAMIANAKIS2023,li2024impact,GAUNT20171,Elmallah_2022,en13195083}. They increase peak loads, can shift the timing of the peaks, and make parts of the grid injection constrained instead of offtake constrained (i.e., the injection peak is higher than the offtake peak). If the electrical load exceeds the capacity of the grid, voltage limits can be violated or thermal capacity can be surpassed. This can slow down or halt the installation of LCTs on the congested parts of the LVG, thereby hindering the transition towards clean energy \cite{iea2023Grids}.

An important hurdle for understanding the behavior of LCTs on the LVG is the lack of datasets of electricity consumption profiles that:
\begin{itemize}
    \item measure actual day-to-day usage patterns of residential customers with LCTs (EV, HP, and PV),
    \item are recently measured (somewhere in the last few years),
    \item are large enough for statistical analysis, including a large enough amount of connections with LCTs.
\end{itemize}
Electricity consumption is measured with a smart (or digital) meter. There are datasets containing smart meter measurements of LCTs, see Ref.~\cite{HABEN2021117798} for a list of open-access datasets of smart meter measurements related to the LVG.
There is, however, a lack of datasets that are simultaneously large, recent, include a significant number of LCTs, and are measured in real-world conditions (instead of, e.g., a trial).
Studies often rely on simulated data \cite{DAMIANAKIS2023}, or a limited set of real measurements which are augmented to create diversity \cite{VELDMAN2013}.
Because there is a large diversity in residential consumption patterns, and consumption is highly stochastic on the level of a single connection, modeling all (or most) of the variation in factors such as occupancy behavior, PV module orientation, and installed devices (pool heating, plug-in PV, \ldots) is difficult.

In this paper, we analyzed a large dataset of 42,089 smart meter profiles from residential LV customers. This dataset:
\begin{itemize}
    \item measures actual day-to-day usage patterns of residential customers with LCTs (EV, HP, and PV),
    \item is recently measured, as all measurements are from 2022,
    \item is large enough for statistical analysis, including a large enough amount of connections with LCTs: 76\% of profiles contain PV, 862 profiles have a HP and 1924 connections have an EV.
\end{itemize}
The dataset was collected by Fluvius cvba, the Distribution System Operator (DSO) of Flanders, Belgium. They were measured at quarter-hour resolution, and reflect real day-to-day consumption patterns in Flanders. These smart meter profiles were coupled with weather data from the ERA5-land dataset \cite{era5}.

To investigate the effect of different LCTs on feeder load, we modeled feeders with these LCTs by creating subsets of the data that, e.g., only contain profiles with EVs. We sampled profiles from these subsets to simulate feeder loads for LV feeders of different sizes, from 10 connections (small LV feeder) to 250 connections (medium-voltage to low-voltage transformer level). We investigated the influence of LCTs on the size and timing (day of year and hour of the day) of the feeder peaks, for both offtake and injection. This sampling procedure is repeated many times, to investigate the full variation in feeder behavior. A visual analysis of the feeder-level load profiles shows that the classical duck curve is replaced by a night-camel curve for feeders with only HPs and a night-dromedary curve for feeders with only EVs charging faster than 6.5 kW. 

Our proposed method is simple to implement, yet returns many important and generic insights that are relevant to a DSO. It can be used to understand the consumption behavior of different types of customers, and shows important qualitative trends of how feeder peaks will change when a large number of LCTs are installed, which is relevant input for DSO grid planners.

There are some limitations to our method, as discussed in detail in Section \ref{sec:discussion}.
Our main simplifying assumption when modeling feeder loads is that the profile dataset is a representative sample from the full population of Flanders. This is likely not the case, as the dataset contains customers that proactively requested higher-resolution smart metering data, and hence is biased towards energy-interested people. The estimation of the change in feeder peak when adding an EV (HP) is added is done by comparing feeders with only EVs (HPs) and without any EVs (HPs), which is also an approximation, since for most feeders not all connections will have an EV (HP) in the future. In this comparison, it is furthermore assumed that the populations with and without an EV (HP) are similar.
Finally, it is important to note that our method investigates \textit{current} consumption profiles. Consumption patterns will likely change in the future. Nevertheless, our method provides a way to keep track of changes in consumption behavior.

To summarize, our main contributions are:
\begin{itemize}
    \item a large, recently measured (2022) dataset that contains actual day-to-day consumption of residential houses at quarter-hour resolution, including those with EVs, HPs, and PV,
    \item an approach to model the consumption on the feeder level, including a collection of insightful visualizations,
    \item quantitative estimates of the peak contribution of EVs and HPs on the feeder peak, which is the relevant quantity for DSOs,
    \item qualitative insights into when the feeder peak occurs, how this depends on the installed LCTs, and what factors play a role in the timing of the feeder peak. The visualizations provide insight into how the consumption patterns depend on the weather (temperature and amount of sunshine) and how the weather influences when the feeder peaks occur.
\end{itemize}

The rest of this paper is organized as follows. Section \ref{sec:relatedlit} describes the related literature. The methods are discussed in Section \ref{sec:methods}. Section \ref{sec:results} describes the results. A discussion of the results, the dataset representativeness and the modeling assumptions is given in Section \ref{sec:discussion}. A conclusion is presented in Section \ref{sec:conclusion}.

\section{Related literature\label{sec:relatedlit}}

\subsection{Modeling consumption profiles}

To estimate the effect of LCTs on the LVG, one has to take into account the diversity and stochasticity of individual residential consumption profiles. The four main methods that tackle this challenge are:
\begin{itemize}
    \item sampling from real data \cite{LOVE2017,barteczko2015after},
    \item using a small set of real data that is synthetically enhanced to create diversity \cite{VELDMAN2013,veldman2011,Bollerslev2022,GONZALEZVENEGAS2021},
    \item physics-based (bottom-up) modeling \cite{RICHARDSON20101878,flett2022_modelica_admd,PROTOPAPADAKI2017268,pflugradt2013analysing},
    \item using generative artificial intelligence models to generate consumption profiles \cite{synthLiang2022,synthGu2019}.
\end{itemize}
Other techniques, such as reference load profiles \cite{energiewandlung2008vdi} (also called standard load profiles) have, by definition, limited capability to capture the diversity of real-world consumption, as they average the consumption of several profiles, thereby removing the stochasticity of real consumption profiles.

Because the interest is in peaks on the level of the distribution grid, the After Diversity Maximum Demand (ADMD) is often investigated \cite{LOVE2017,barteczko2015after,flett2022_modelica_admd}. It is defined as the maximum demand at an aggregated level, i.e., the maximum after accounting for the diversity in when the peaks of the individual demand profiles occur. It is a more general concept than feeder peak, which is the word we use. ADMD doesn't necessarily refer to a feeder, but can refer to any collection of customers. It can also refer to, e.g., a population of only heat pumps or EVs instead of all consumption on a connection. Because we are directly modeling the feeder peaks, we mainly use the word feeder peak.

\subsection{Results using real consumption profiles}

We now discuss the results obtained by using real profiles in more detail, as this is most closely related to our work.

\subsubsection{Heat pumps}

A dataset of heat pumps from the UK found 1.7 kW of ADMD for the heat pumps themselves (without taking into account the other consumption on the feeder) \cite{LOVE2017}. The ADMD of EVs and heat pumps separate from the rest of consumption for a dataset from the UK \cite{ukdataset} is reported in \cite{barteczko2015after}. The ADMD of the total consumption profiles, containing both EV or HP consumption combined with all other consumption on the feeder, is also reported, but not analyzed in detail. In Ref.~\cite{WANG2020116780}, the ADMD for connections with a heat pump on very cold days was found to be between 1 kW and 2 kW.
The ADMD for the air-source heat pumps themselves, for 20 homes in Ireland, for a very cold day was found to be 1.5 kW (from 2.3 to 3.8 kW) \cite{irish_ashp_2021}.

\subsubsection{EVs}

For EVs, consumption profiles are often estimated by combining real EV charging profiles with models for the arrival times and required charging energy \cite{Bollerslev2022,GONZALEZVENEGAS2021,SILBER2024,Fani2023,hungbo2023impact,9667519}. Data on arrival times or length of charging sessions is taken from real data in field trails or modeled from, e.g., travel data. We are not aware of articles that directly sample real-world residential EV data, without an intermediate modeling step. Because we don't add EV charging profiles on top of other consumption data, we don't need to create a statistical model of EV charging. Instead, we investigated smart meter data of connections where EV charging has been detected. Because we have a large dataset, we assume that the real variety in charging behavior (start time, length of charging session, \ldots) is captured in the real smart meter data.

\subsection{Output variables}

There are several relevant output variables that are investigated in the literature. One can estimate how much the feeder peak increases when adding an LCT to the grid. More specifically, one can estimate the contribution to the feeder peak of a device, in the form of ADMD \cite{LOVE2017,admd_probabilistic_2019} or simultaneity factor \cite{Bollerslev2022}. Because the actual outcome of interest is grid congestion, consumption profiles are often used to run power flows on real or synthetic LVGs, to estimate the likelihood of voltage violations or transformer- or line overloading \cite{DAMIANAKIS2023,hungbo2023impact,9667519}.

We do not perform power flow simulations on a set of representative feeders. We present new ways to analyze and visualize results from a random (Monte Carlo) profile sampling approach, as discussed in the results, Section \ref{sec:results}. Besides a quantitative assessment of how much LCTs add to the feeder peak, several qualitative insight on the impact of LCTs on the LVG are provided. Compared to similar articles, we do a more extensive analysis of when the feeder peak occurs: on what hour and day of the year, how does the timing shift when LCTs are added, and how this is related temperature and amount of sunshine.

\section{Materials and Methods\label{sec:methods}}

\subsection{Profile and weather data\label{sec:methods_sampling}}

Before the data was investigated, all smart meter data was fully anonymized by Fluvius. Electricity consumption is measured every 15 minutes, for a whole year. The analysis was performed in local Belgian time. If a day misses an hour because of daylight savings time, that hour was imputed by the values of the previous hour. If there was an extra hour because of daylight savings time, the redundant hour was removed. Because this influences two hours in a full year, these changes are expected to have a negligible influence on the end results. Labels for the presence of an EV charging point and heat pump were provided by Fluvius.

Weather variables at hourly resolution were obtained from the ERA5-Land dataset \cite{era5}. The spatial resolution is 9 by 9 kilometers. Temperature of air at 2 meters above the surface in degree Celsius and surface solar radiation downwards (ssrd) in units of $kW / m^2$ are included in the analysis. A geographical region in the middle of Flanders, namely the region containing the city of Mechelen, was selected for the weather data. This was done because, although the smart meter data was collected over the whole of Flanders, all profiles are analyzed together.

\subsection{Feeder modeling with profile sampling\label{sec:samplingmethod}}

\begin{figure}
\centering
\includegraphics[width=0.8\textwidth]{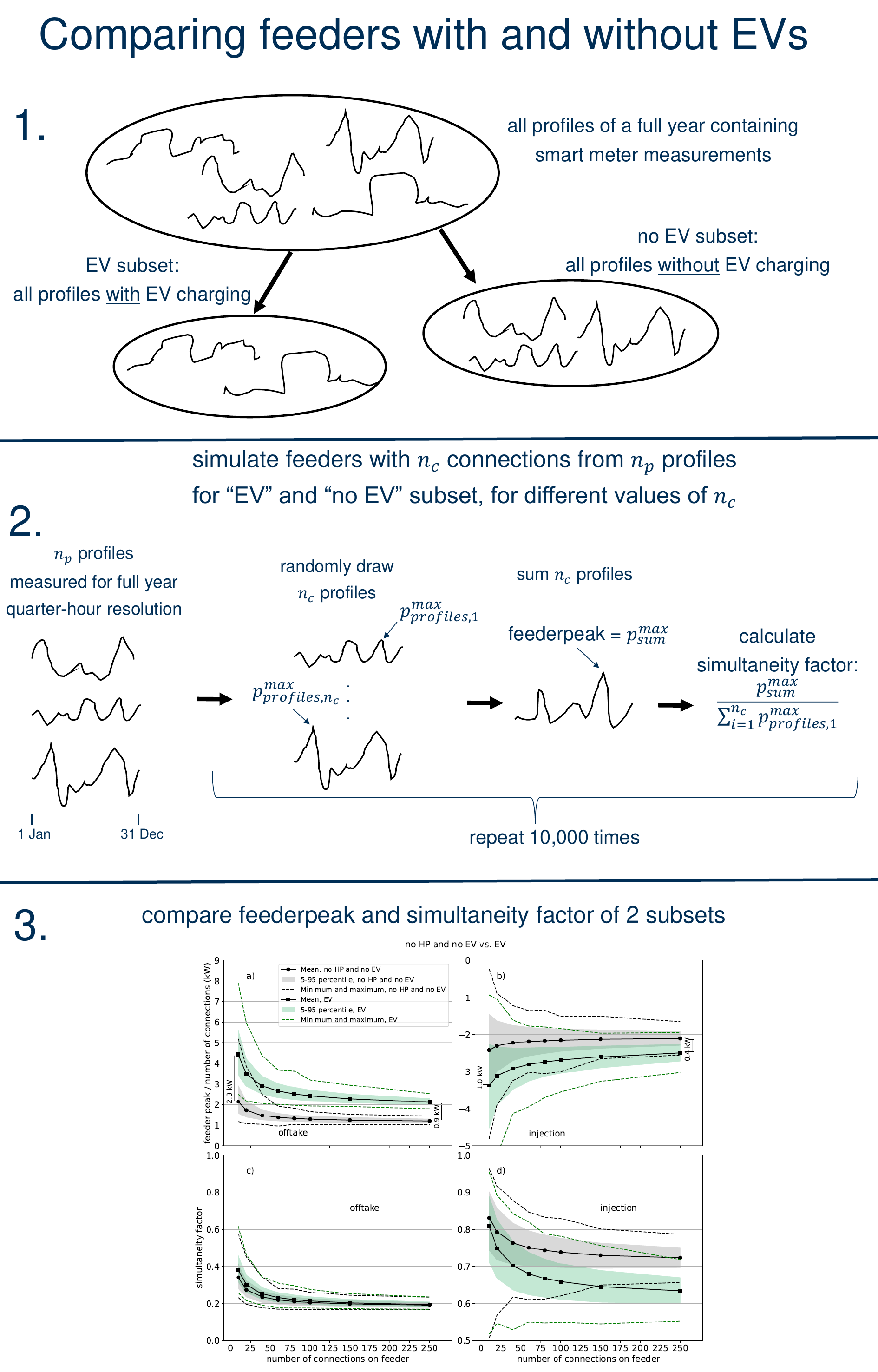}
\caption{Visualization of the main steps of our approach, for the case of comparing feeders with and without EVs. Step 1: all smart meter measurements are divided into subsets: connections with and without EV. Step 2: feeders are simulated for each subset, by summing the power consumption of a sample of profiles, and calculating the feeder peak and simultaneity factor. Step 3: the effect of adding EVs to LV feeders can be investigated by comparing the results for the different subsets.}
\label{fig:drawing_method}
\end{figure}

The distribution of the LV-feeder load was estimated as follows.
\begin{enumerate}
    \item Consider one of the subsets of the profile dataset (e.g.,~all connections containing an EV). Take $n_p$ the number of profiles in the dataset, $n_c$ the number of connections of the feeder, and $n_q$ the number of quarter hours in a full year. Denote $\mathbf{P}$ as the matrix of smart meter power measurements (kW) of size $n_p \times n_q$.
    \item Randomly draw $n_c$ profiles (i.e, rows) from $\mathbf{P}$. These profiles represent a LV feeder. They are represented by a matrix $\mathbf{P}_{feeder}$ of size $n_c \times n_q$.
    \item Calculate $\mathbf{p}^{max}_{profiles}$ of size $n_c \times 1$: maximum power (quarter-hour resolution) of each profile in $\mathbf{P}_{feeder}$
    \item Calculate the vector $\mathbf{p}^{sum}_{profiles}$ of size $1 \times n_q$: the sum of individual power consumptions of all profiles in $\mathbf{P}_{feeder}$, for each quarter hour. Calculate the feeder peak:
    \begin{equation}\label{eq:feederpeak}
    p^{max}_{sum}\textrm{: maximum power in } \mathbf{p}^{sum}_{profiles},
    \end{equation}
    and calculate:
    \begin{equation}\label{eq:simfactor}
    \textrm{simultaneity factor}  = p^{max}_{sum} / \left( \sum_{i=1}^{n_c} p^{max}_{profiles, i} \right).         
    \end{equation}
    Store the time at which $p^{max}_{sum}$ occurs.
\end{enumerate}
 Repeat this procedure many times (in our case, 10,000), each time with a different set of randomly drawn profiles. Pseudo code is given in Algorithm \ref{alg:mcsampling}. A drawing of the proposed method is shown in Figure \ref{fig:drawing_method}, for the case of comparing feeders with or without EVs. The same procedure is followed for injection peaks, where the minimum power instead of the maximum power is calculated. 

There are articles that use the same method, such as \cite{LOVE2017} for calculating the ADMD of heat pumps. A related method is probabilistic fitting instead of sampling \cite{admd_probablistic_2004,admd_probabilistic_2019}, in order to avoid having to sample all possible combinations of connections. However, we draw enough samples (10,000) to render this step unnecessary.

\begin{algorithm}
\caption{Profile sampling to model feeder load}
\begin{algorithmic}[1] 
\STATE $n_p$: number of profiles in dataset
\STATE $n_c$: number of connections on feeder
\STATE $n_q$: number of quarter hours in a full year
\STATE $\mathbf{P}$: matrix of smart meter power measurements (kW), size $n_p \times n_q$
\STATE $n_s$: number of samples = 10,000
\FOR{i = 1:$n_s$}
    \STATE create matrix $\mathbf{P}_{feeder}$ of size $n_c \times n_q$: draw $n_c$ random profiles from $\mathbf{P}$
    \STATE calculate vector $\mathbf{p}^{max}_{profiles}$ of size $n_c \times 1$: maximum power (quarter-hour resolution) of each profile in $\mathbf{P}_{feeder}$ 
    \STATE calculate vector $\mathbf{p}^{sum}_{profiles}$ of size $1 \times n_q$: sum of individual profiles in $\mathbf{P}_{feeder}$, for each quarter hour 
    \STATE calculate scalar $p^{max}_{sum}$: maximum power in $\mathbf{p}^{sum}_{profiles}$
    \STATE calculate simultaneity factor $= p^{max}_{sum} / \left( \sum_{i=1}^{n_c} p^{max}_{profiles, i} \right)$
    \STATE store time at which $p^{max}_{sum}$ occurs
    \STATE save results
\ENDFOR
\end{algorithmic}
\label{alg:mcsampling}
\end{algorithm}

\subsection{Computational details}

All code was written in Python \cite{python_book}. To speed up the profile sampling, the code ran on a 96-core Linux server. Time-critical parts of the code were written in Numpy \cite{harris2020array} and Polars. The computation time for all the presented results was less than 24 hours. The random seed was fixed, so the results are reproducible. This method has been shown to scale to a dataset containing approximately 200,000 profiles, which is an order of magnitude larger than the one investigated in this manuscript. However, we expect that more specialized code or hardware would be needed to make this work for datasets containing millions of profiles or more.

\section{Results\label{sec:results}}

\subsection{Data properties}

We defined four subsets of the full dataset:
\begin{itemize}
\item \textit{HP}: the subset of all profiles with a HP. 
\item \textit{EV}: the subset of all profiles with an EV. 
\item \textit{EV, high power}: the subset of all profiles with an EV that has a maximum charging power higher than 6.5 kW.
\item \textit{no HP, no EV}: the subset of all profiles without a HP and without an EV.
\end{itemize}
The subsets \textit{EV} and \textit{HP} have 67 connections in common, and \textit{EV, high power} and \textit{HP} have 30 connections in common.

\begin{table*}[hbt]
\caption{Properties of the subsets of the full dataset. Mean and standard deviation (between parentheses) are provided. PV stands for PV inverter power (kVA). Conn.~power stands for connection power (kVA) and Consumption stands for net consumption over a full year (kWh).}
\label{tab:dataset}
\begin{tabular}{@{}lcccc}
\hline
                  & no HP, no EV & HP        & EV         & EV, high power \\
\hline
Number            & 39,303        & 862       & 1924       & 758            \\
\% PV             & 75           & 94        & 86         & 85             \\
PV  & 4.3 (1.7)    & 5.6 (2.1) & 5.4 (2.1)  & 5.6 (2.2)      \\
Conn.~power & 14.6 (6.3) & 19.9 (5.5) & 18.6 (6.8) & 20.6 (6.4) \\
Consumption & 900 (2784) & 1646 (3631) & 4030 (3944) & 4942 (3996) \\
\hline
\end{tabular}
\end{table*}

\begin{figure}
\centering
\includegraphics[width=0.75\textwidth]{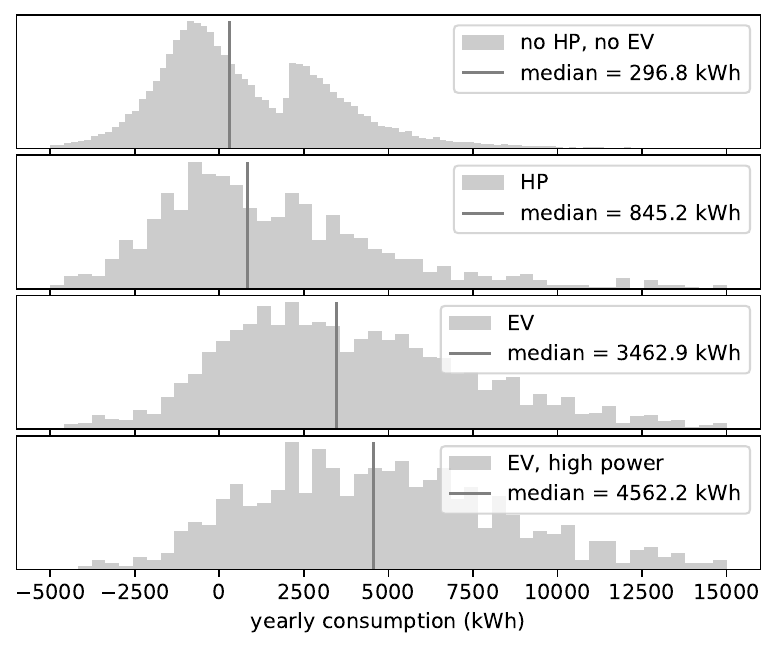}
\caption{Histogram of the yearly consumption of all profiles in each subset. The subsets are shown in the legend, and the median is also shown.}
\label{fig:histo_consumptionprofiles}
\end{figure}

\begin{figure}
\centering
\includegraphics[width=0.75\textwidth]{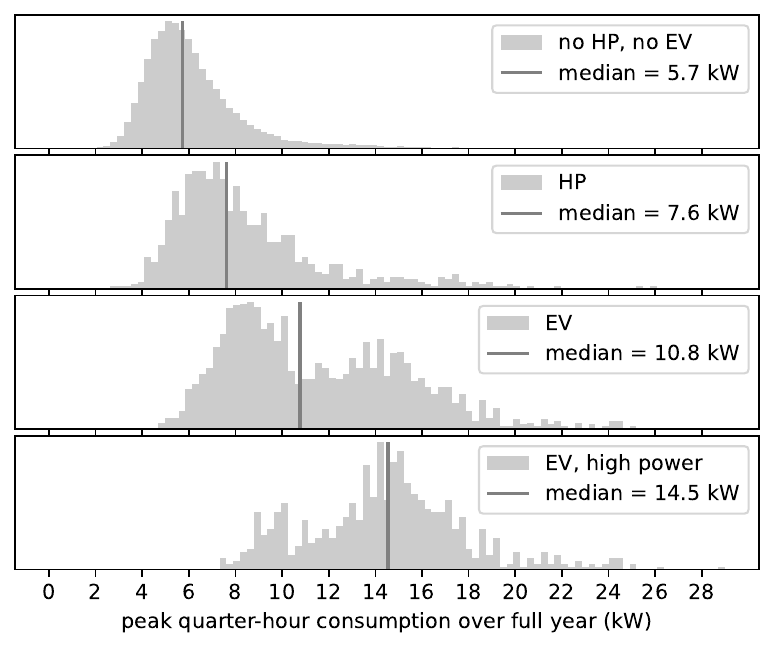}
\caption{Histogram of the quarter-hour with the highest consumption (i.e., the consumption peak) over the full year, for all profiles in each subset. The subsets are shown in the legend, and the median is also shown.}
\label{fig:histo_peakprofiles}
\end{figure}

A summary of the properties of the dataset can be found in Table \ref{tab:dataset}. We see that connections with an EV or HP have on average more PV and a stronger connection power. The net yearly consumption is low for the \textit{no HP, no EV} category because the PV installations lead to negative yearly consumption for many connections. This can be seen in Figure \ref{fig:histo_consumptionprofiles}, where the distribution of yearly consumption is shown for the 4 subsets. A clear bi-modal distribution is observed for \textit{no HP, no EV}, with a local peak around 3000 kWh, which is the typical yearly consumption for residential users in Flanders (without PV), and a second peak of negative yearly consumption for connections with PV. 
Histograms of the quarter-hour with the highest consumption (i.e., peaks of the individual profiles) for the subsets are shown in Figure \ref{fig:histo_peakprofiles}. Connections with a HP have higher peaks (on average), and connections with an EV have higher peaks still. This is as expected. However, the feeder peaks $p^{max}_{sum}$ are determined by the individual peaks and the simultaneity factor, see Section \ref{sec:samplingmethod} and Figure \ref{fig:drawing_method}. 

The dataset is for customers that proactively requested higher-resolution smart metering data, i.e., quarter hour measurements rather than daily measurement values. This means that the dataset is biased towards people interested in their energy consumption, which are typically also more interested in the energy transition compared to the average Flemish citizen. It is therefore important to emphasize that our results do not represent current consumption patterns on the grid. This bias is reflected in the higher than average PV ownership in the dataset, which for the whole Flemish population is 18\%. For the whole Flemish population the average installed PV inverter power is 4.6 kVA and the average connection power is 14.5 kVA, which are similar numbers compared to the subset \textit{no HP, no EV}. Our results provide insight into consumption behavior on the LVG when a larger amount LCTs are installed. As discussed in Section \ref{sec:datarep}, this dataset is not an exact representation of consumption behavior in the future. 

\subsection{Load profile behavior}

\begin{figure}
\includegraphics[width=\textwidth]{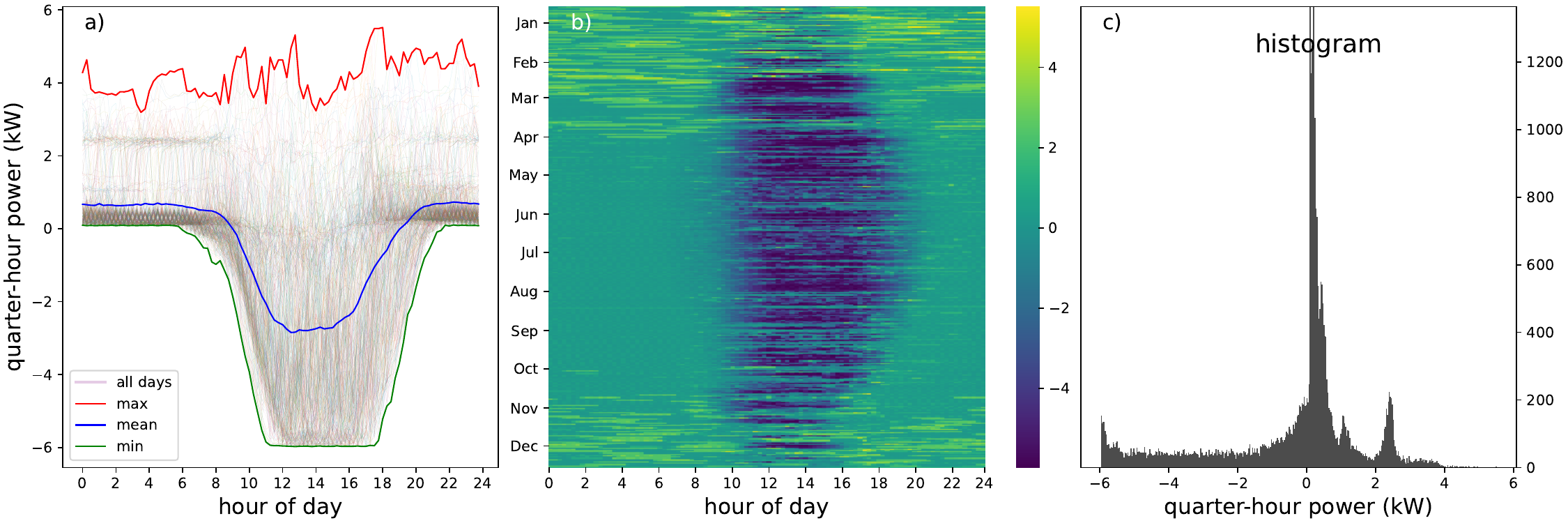}
\caption{Full year of quarter-hour electricity consumption data (kW) for a connection with a heat pump. a) Time series for each day of the year, plotted together. The mean (blue), maximum (red) and minimum (green) value for each quarter hour of the year are also shown. b) Heat map of consumption time series (kW). Hour of day is on the x-axis and day of the year is on the y-axis. c) Histogram of quarter-hour power values. 
The heat pump shows modulating consumption, mostly at night during winter. Peak power of the heat pump is around 2.5 kW, as can be seen from the histogram and the profile plots. The power of the PV inverter is approximately 6 kVA.}
\label{fig:profile_hp}
\end{figure}

\begin{figure}
\includegraphics[width=\textwidth]{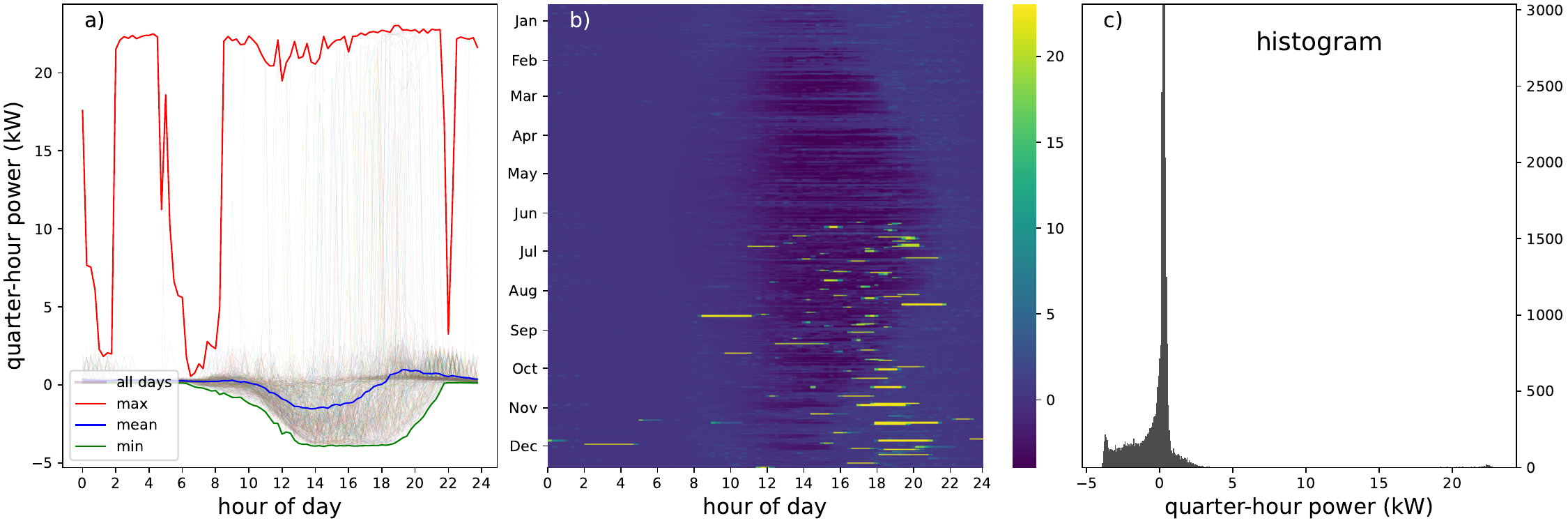}
\caption{Full year of quarter-hour electricity consumption data (kW) for a connection with an EV. 
a) Time series for each day of the year, plotted together. The mean (blue), maximum (red) and minimum (green) value for each quarter hour of the year are also shown. b) Heat map of consumption time series (kW). Hour of day is on the x-axis and day of the year is on the y-axis. c) Histogram of quarter-hour power values. 
Right: histogram of quarter-hour power values. Charging occurs at 22 kW, mostly in the evening between 16h and 22h. PV inverter power is around 4.5 kVA.}
\label{fig:profile_ev}
\end{figure}

\begin{figure}
\includegraphics[width=\textwidth]{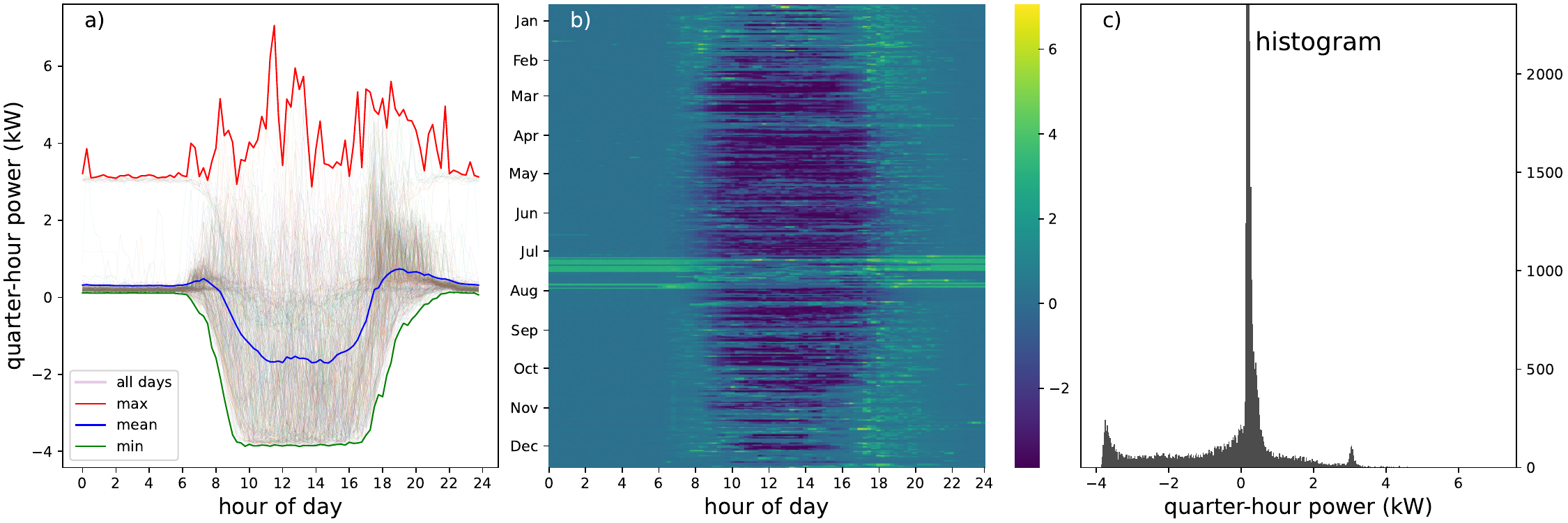}
\caption{Full year of quarter-hour electricity consumption data (kW) for a connection with an under-dimensioned PV inverter. 
a) Time series for each day of the year, plotted together. The mean (blue), maximum (red) and minimum (green) value for each quarter hour of the year are also shown. b) Heat map of consumption time series (kW). Hour of day is on the x-axis and day of the year is on the y-axis. c) Histogram of quarter-hour power values. 
High and constant consumption can be observed during the summer months.}
\label{fig:profile_pv_underdim}
\end{figure}

\begin{figure}
\includegraphics[width=\textwidth]{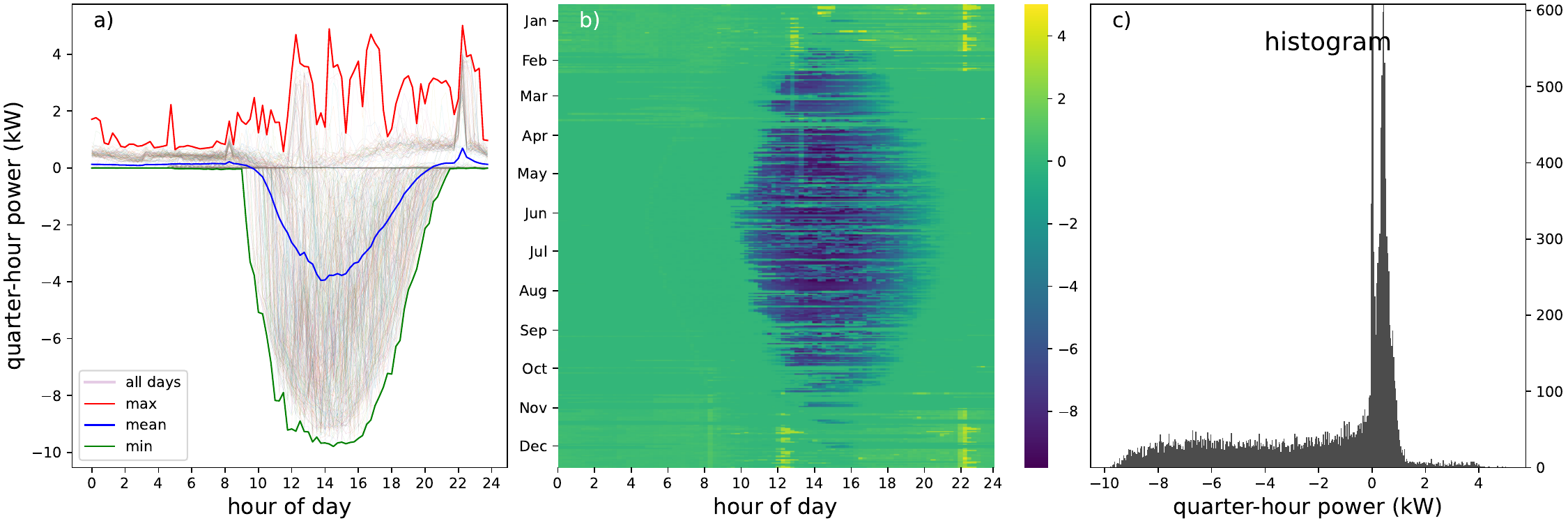}
\caption{Full year of quarter-hour electricity consumption data (kW) for a connection with a 10 kVA PV installation. 
a) Time series for each day of the year, plotted together. The mean (blue), maximum (red) and minimum (green) value for each quarter hour of the year are also shown. b) Heat map of consumption time series (kW). Hour of day is on the x-axis and day of the year is on the y-axis. c) Histogram of quarter-hour power values. 
There is likely a battery installed, since there is no consumption at all at night during summer and parts of spring and autumn. Noticed the timed consumption peaks around 12h and 22h.}
\label{fig:profile_battery}
\end{figure}

In this section, we plot some of the smart meter profiles in the dataset, see Figures \ref{fig:profile_hp}-\ref{fig:profile_battery}. These profiles were chosen to highlight common consumption patterns observed in the dataset and to show some of the discovered peculiarities. Figures \ref{fig:profile_hp}-\ref{fig:profile_battery} show a full year of quarter-hour consumption data in units of kW. They consist of three plots. On the left, the consumption time series for each day are plotted on top of each other. The minimum, mean, and maximum over the whole year, for each quarter hour, are shown as well. In the middle, a heat map of the consumption data is shown, with the hour of the day on the x-axis, and the day of the year on the y-axis. On the right, a histogram of all quarter-hourly power values is plotted.

Figure \ref{fig:profile_hp} shows a connection with a heat pump. We can see the modulating behavior of the heat pump in the left and middle sub-figures. Many heat pumps show this modulating power consumption behavior, which is expected \cite{BRUDERMUELLER2023}; others have more constant consumption profiles. There is almost no heat pump activity in the summer months. PV production is visible for all months: also in January and December there are moments of power injection. In the histogram on the right sub-figure, we see that the heat pump operates at a power of about 2.5 kW, as there is a local peak in the histogram around that value. The power of the PV inverter is approximately 6 kVA, as the injection from PV is limited at around 6 kW.

A connection with an EV is shown in Figure \ref{fig:profile_ev}. The EV is charged at 22 kW, mostly in the evening between 16h and 22h.  The installed PV has little effect on peak offtake power, despite this being an installation with an inverter of approximately 4.5 kVA. EV charging only start in June, so most likely the EV charging point was installed in the middle of the year.

PV inverter power was often observed to be smaller than peak installed PV power, i.e., the inverters are undersized. Examples are shown in Figures \ref{fig:profile_hp} and \ref{fig:profile_pv_underdim}. For Figure \ref{fig:profile_pv_underdim}, maximal injection power is around 4 kW, which is reached quite early in the day (around 9h). This means that the inverter capacity is significantly smaller than the installed PV capacity. There is some unusually high and constant consumption in summer (July and August). This could, for example, be swimming pool heating or air conditioning. This is one example of many where consumption patterns are observed for which the underlying reasons are uncertain.

What is likely a connection with a battery is shown in Figure \ref{fig:profile_battery}. There is almost never offtake between April and September, only injection during the middle of the day and zero offtake at night. This is likely because a battery is supplying all the energy needed to cover the consumption demand at night. Note the timed consumption peaks around 12h and 22h. In the dataset, there are many connections with timed consumption peaks.

We believe it would be difficult to capture all of the observed behavior in a bottom-up modeling approach. There are a large amount of unexpected consumption patterns, which are sometimes even for experts difficult to interpret. Guessing the occurrence and distribution of all these phenomena is challenging. For example, we found a lot of consumption peaks that are clearly timed, as in Figure \ref{fig:profile_battery}, but the size and timing of these peaks differs significantly among the profiles. The same holds true for EV charging behavior: the frequency of charging events and their timing during the day varies significantly, making it challenging to model this consumption behavior. Not having to model all this stochasticity is an advantage of our approach, which consists of using a large dataset that measures actual day-to-day consumption. 

\subsection{Feeder peaks and simultaneity factors}

We discuss the results for the feeder peaks and simultaneity factors, and the effect of EVs and HPs on these quantities.

\subsubsection{Peaks and simultaneity factor for feeders\label{sec:admd_ksim_ind_feeders}}

\begin{figure}
\centering
\includegraphics[width=0.95\textwidth]{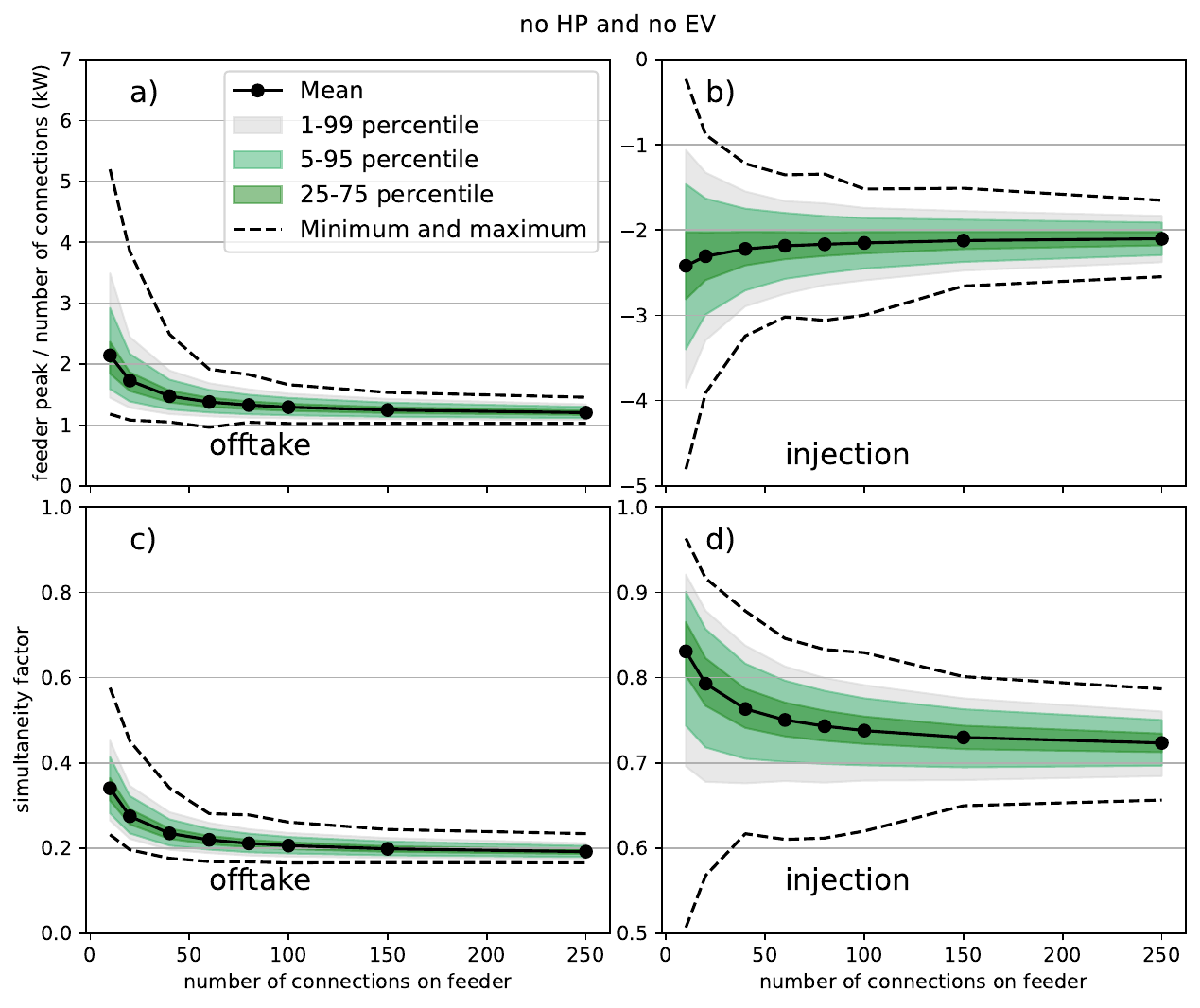}
\caption{Feeder peak per connection (kW) (Eq.~\eqref{eq:feederpeak}) for offtake a) and injection b). Simultaneity factor (Eq.~\eqref{eq:simfactor}) for offtake c) and injection d), as a function of the number of connections on the feeder. For a feeder without heat pumps and without EVs.}
\label{fig:admd_no_hp_no_ev}
\end{figure}

\begin{figure}
\centering
\includegraphics[width=0.95\textwidth]{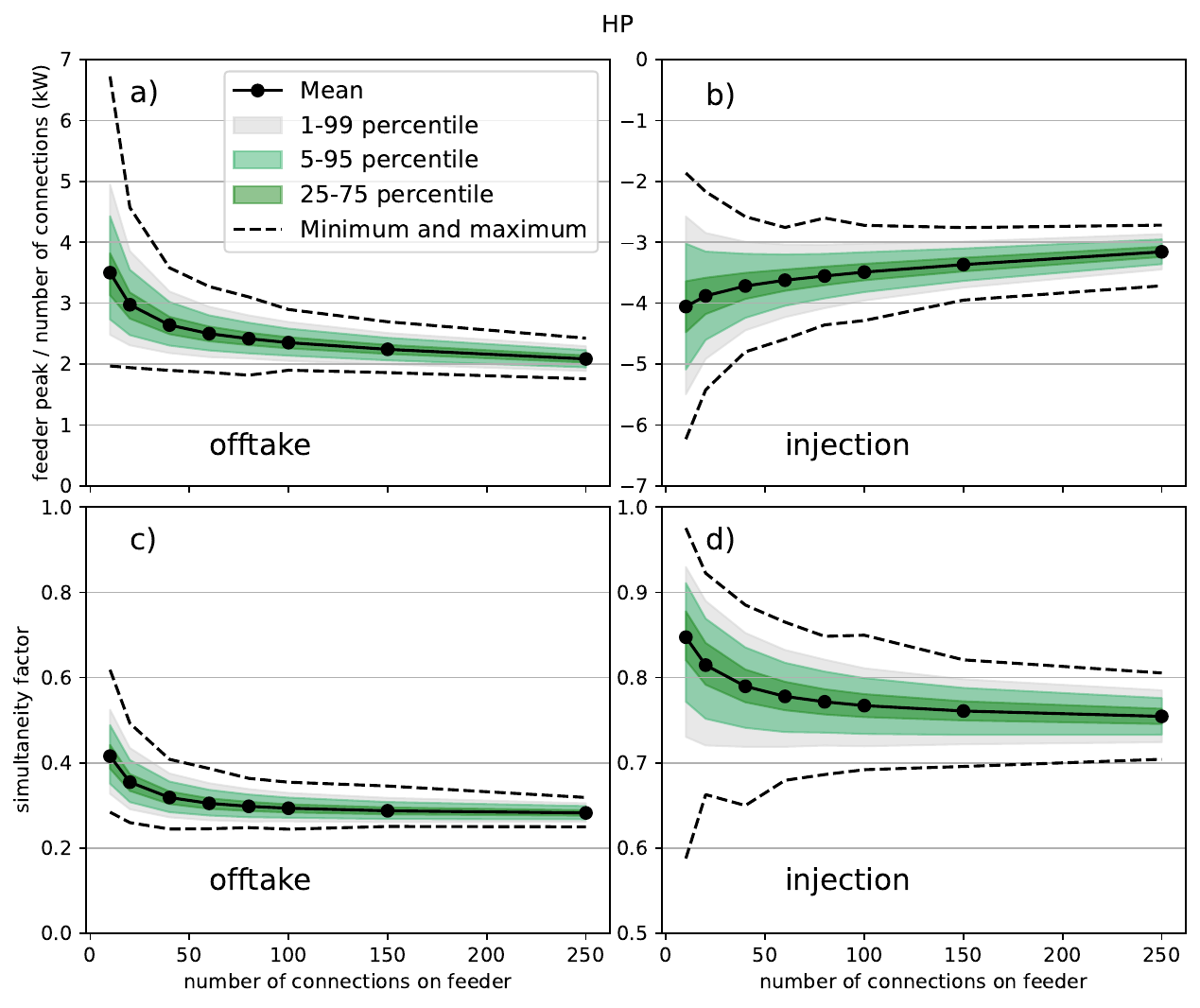}
\caption{Feeder peak per connection (kW) (Eq.~\eqref{eq:feederpeak}) for offtake a) and injection b). Simultaneity factor (Eq.~\eqref{eq:simfactor}) for offtake c) and injection d), as a function of the number of connections on the feeder. For a feeder with only heat pumps.}
\label{fig:admd_ksim_hp}
\end{figure}

\begin{figure}
\centering
\includegraphics[width=0.95\textwidth]{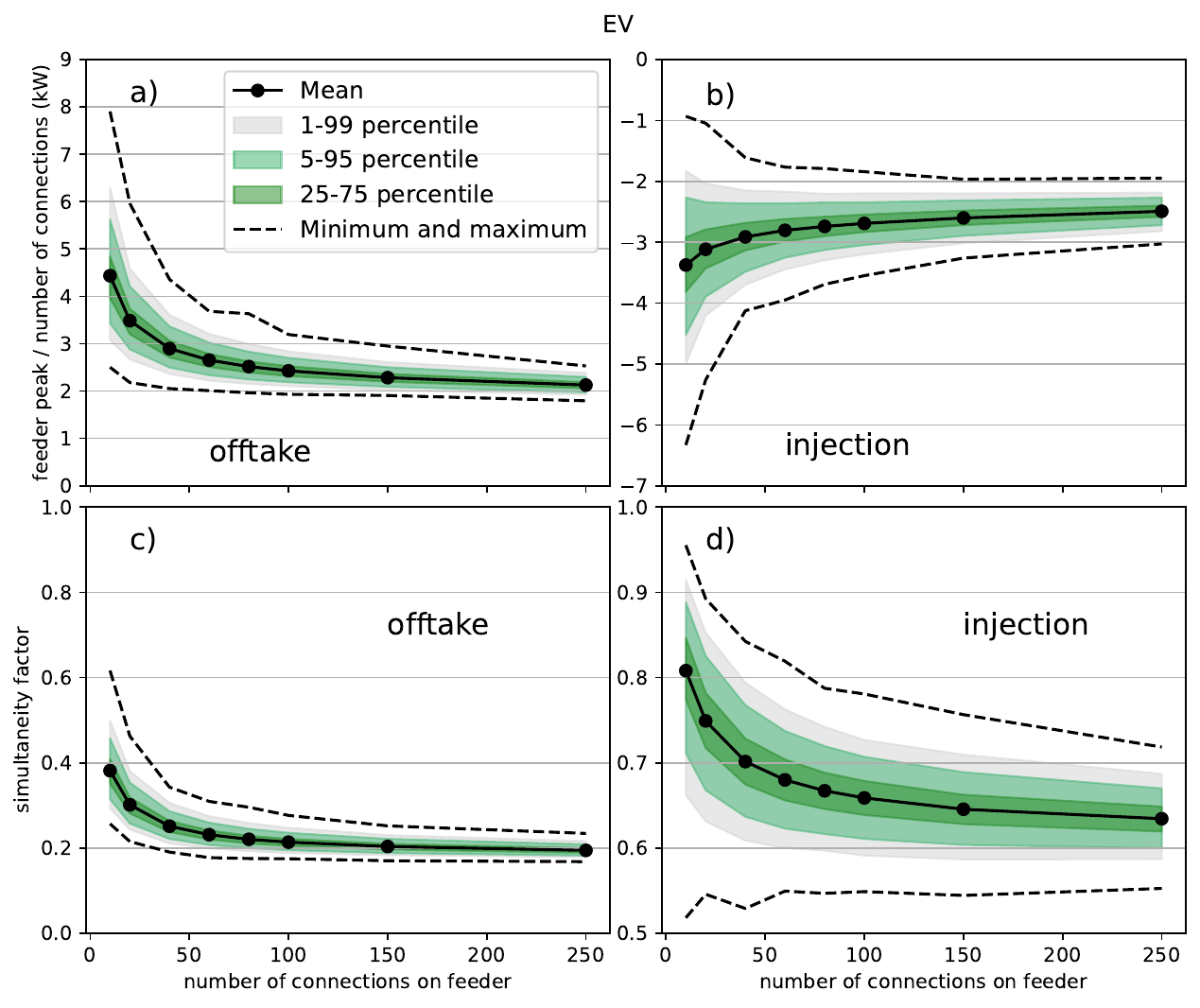}
\caption{Feeder peak per connection (kW) (Eq.~\eqref{eq:feederpeak}) for offtake a) and injection b). Simultaneity factor (Eq.~\eqref{eq:simfactor}) for offtake c) and injection d), as a function of the number of connections on the feeder. For a feeder with only EVs.}
\label{fig:admd_ksim_ev}
\end{figure}

\begin{figure}
\centering
\includegraphics[width=0.95\textwidth]{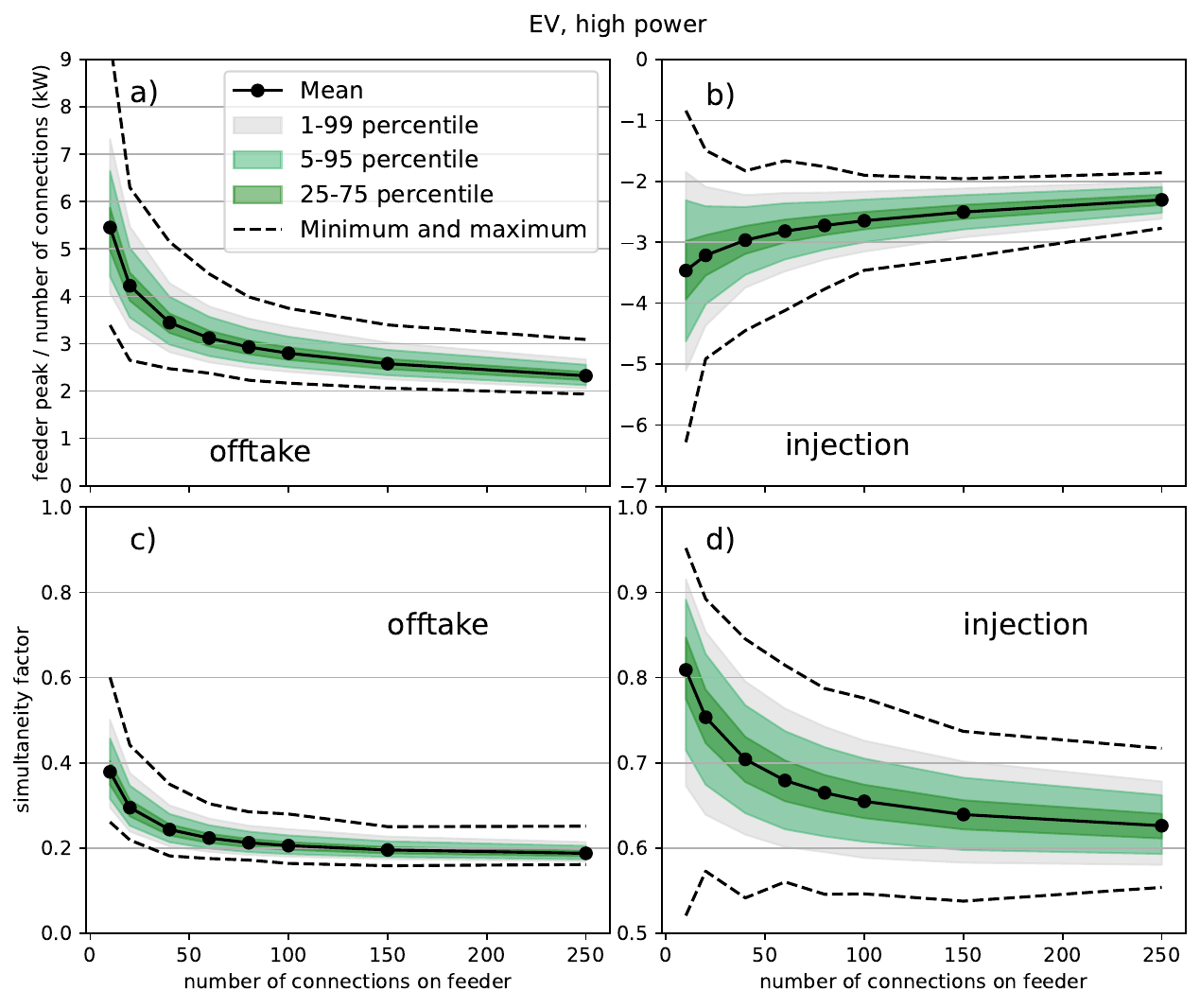}
\caption{Feeder peak per connection (kW) (Eq.~\eqref{eq:feederpeak}) for offtake a) and injection b). Simultaneity factor (Eq.~\eqref{eq:simfactor}) for offtake c) and injection d), as a function of the number of connections on the feeder. For a feeder with only EVs charging at higher power.}
\label{fig:admd_ksim_evhigh}
\end{figure}

The results for \textit{no HP, no EV}, for different number of connections, are shown in Figure \ref{fig:admd_no_hp_no_ev}. 
The feeder peak divided by the number of connections is shown in Figure \ref{fig:admd_no_hp_no_ev}a. The feeder peak is the quarter hour in the year with the highest power demand, over a whole year. It is divided by the number of connections to make the values for different feeder sizes comparable. It also indicates how much grid capacity must be provided per household. Information on how the values are distributed is provided by plotting the $5^{th}$, $25^{th}$, $75^{th}$, and $95^{th}$ percentiles, and the minimum and maximum.
The simultaneity factor is shown in figure \ref{fig:admd_no_hp_no_ev}b. Its distribution is visualized in the same way. The results for the feeder peak and simultaneity factor for injection are shown in, respectively, Figures \ref{fig:admd_no_hp_no_ev}c and \ref{fig:admd_no_hp_no_ev}d.
The same results are shown for \textit{HP} in Figure \ref{fig:admd_ksim_hp}, for \textit{EV} in Figure \ref{fig:admd_ksim_ev}, and \textit{EV, high power} in Figure \ref{fig:admd_ksim_evhigh}.

The following common behavior is observed for all the subsets. The size of the peaks is strongly dependent on the number of connections. We see a convergence of the feeder peak at around 150 connections, which is a typical size for a medium-voltage to low-voltage transformer. Except for EVs with high charging power, Figure \ref{fig:admd_ksim_evhigh}, where the feeder (or transformer) peak is still decreasing from 150 to 250 connections. The spread of feeder-peak values is much higher for a small number of connections compared to a large number of connections. The variation keeps decreasing up until 250 connections, although the decrease is small from 150 to 250 connections.

For the simultaneity factor, a strong decrease is observed with the number of connections. For HPs, Figure \ref{fig:admd_ksim_hp}, we start from approximately 0.4 for 10 connections to 0.25 for 250 connections. The spread of values also strongly decreases as a function of the number of connections. The average simultaneity factor converges at around 100 connections. The \textit{EV} (Figure \ref{fig:admd_ksim_ev}) and \textit{EV, high power} (Figure \ref{fig:admd_ksim_evhigh}) subsets have a simultaneity factor of around 0.4 for 10 connections and converge to around 0.2 for 250 connections. For the \textit{no HP, no EV} subset (Figure \ref{fig:admd_no_hp_no_ev}), the simultaneity factor at 10 connections is 0.35, which is a bit lower compared to the other subsets. It converges to 0.2 for a high number of connections, similar to \textit{EV} and \textit{EV, high power} and a bit lower than \textit{HP}, which is 0.25.
 
For \textit{HP} feeders with 10 connections, the maximum feeder peak is $\approx$ 6.5 kW, which is almost double the average of 3.5 kW and still significantly higher than the $95^{th}$ percentile value of 4.5 kW. This is because of natural variation in the consumption profiles.
Even for 250 connections there is a difference of around 0.5 kW between the minimum and maximum. The average feeder peak decreases significantly when going to 250 connections: from 3.5 kW for 10 connections to 2.0 kW for 250 connections. The same behavior is observed for the other subsets. This shows that stochasticity is an important factor on the LVG; only investigating average values provides insufficient information to estimate the required grid capacity. 

For small feeders, the peaks are the highest for \textit{EV, high power}, followed by \textit{EV} and then \textit{HP}, with \textit{no HP, no EV} being the smallest. This is the expected behavior. At the medium-voltage level (250 connections), the feeder peak per connection is around 2 kW for \textit{EV}, \textit{HP}, and \textit{EV, high power}, while it is only 1 kW for \textit{no HP, no EV}. 

For the subsets \textit{no HP, no EV} and \textit{HP}, the feeders have higher injection peaks than offtake peaks, for all feeder sizes. There is however a high overlap in their distributions. For \textit{EV} and \textit{EV, high power}, offtake peaks are higher than injection peaks for small feeders, but have similar values for large feeders.

\subsubsection{Contribution of HP and EV to feeder peak and simultaneity factor}

To estimate how much a HP and an EV add to the feeder peak, and how they change the simultaneity factor on the feeder peak, we compared the feeder peak per connection of feeders without any HPs (EVs) with feeders with only HPs (EVs). The difference between the two is an approximation of the contribution of a single HP (EV) to a feeder.

\begin{figure}
\centering
\includegraphics[width=1.\textwidth]{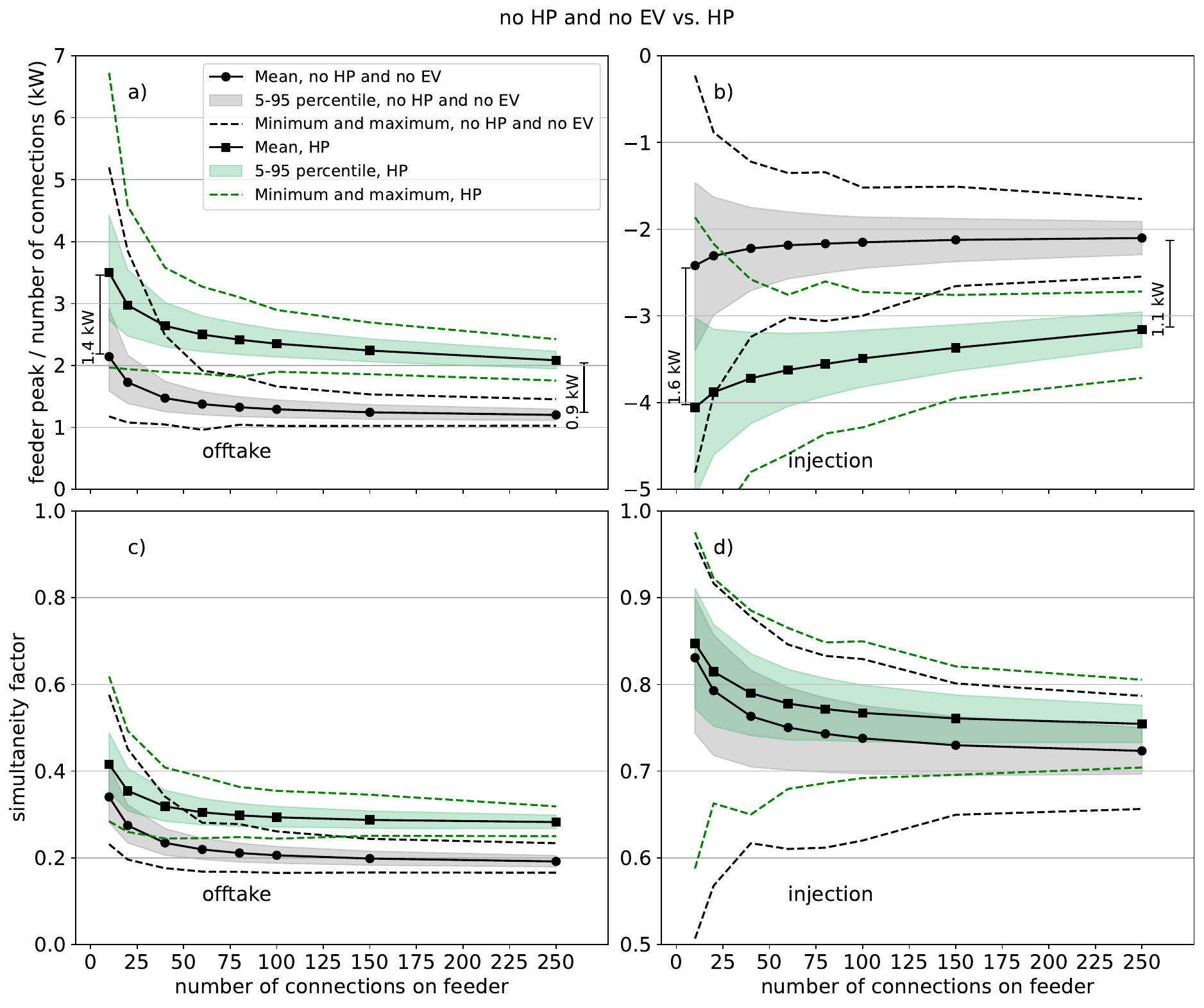}
\caption{Feeder peak per connection (kW)  (Eq.~\eqref{eq:feederpeak}) and simultaneity factor (Eq.~\eqref{eq:simfactor}) for \textit{no HP, no EV} feeders versus \textit{HP} feeders, for different number of connections, both offtake and injection.}
\label{fig:hp_add}
\end{figure}

The results for heat pumps are shown in Figure \ref{fig:hp_add}.
A single heat pump adds between 1.4 kW (10 connections) and 0.9 kW (250 connections) to the feeder peak, as shown in Figure \ref{fig:hp_add}a. The change in injection peak is shown in Figure \ref{fig:hp_add}b.
The change in simultaneity factor when adding a HP is shown in Figure \ref{fig:hp_add}c for offtake and Figure \ref{fig:hp_add}d for injection.

\begin{figure}
\centering
\includegraphics[width=1.\textwidth]{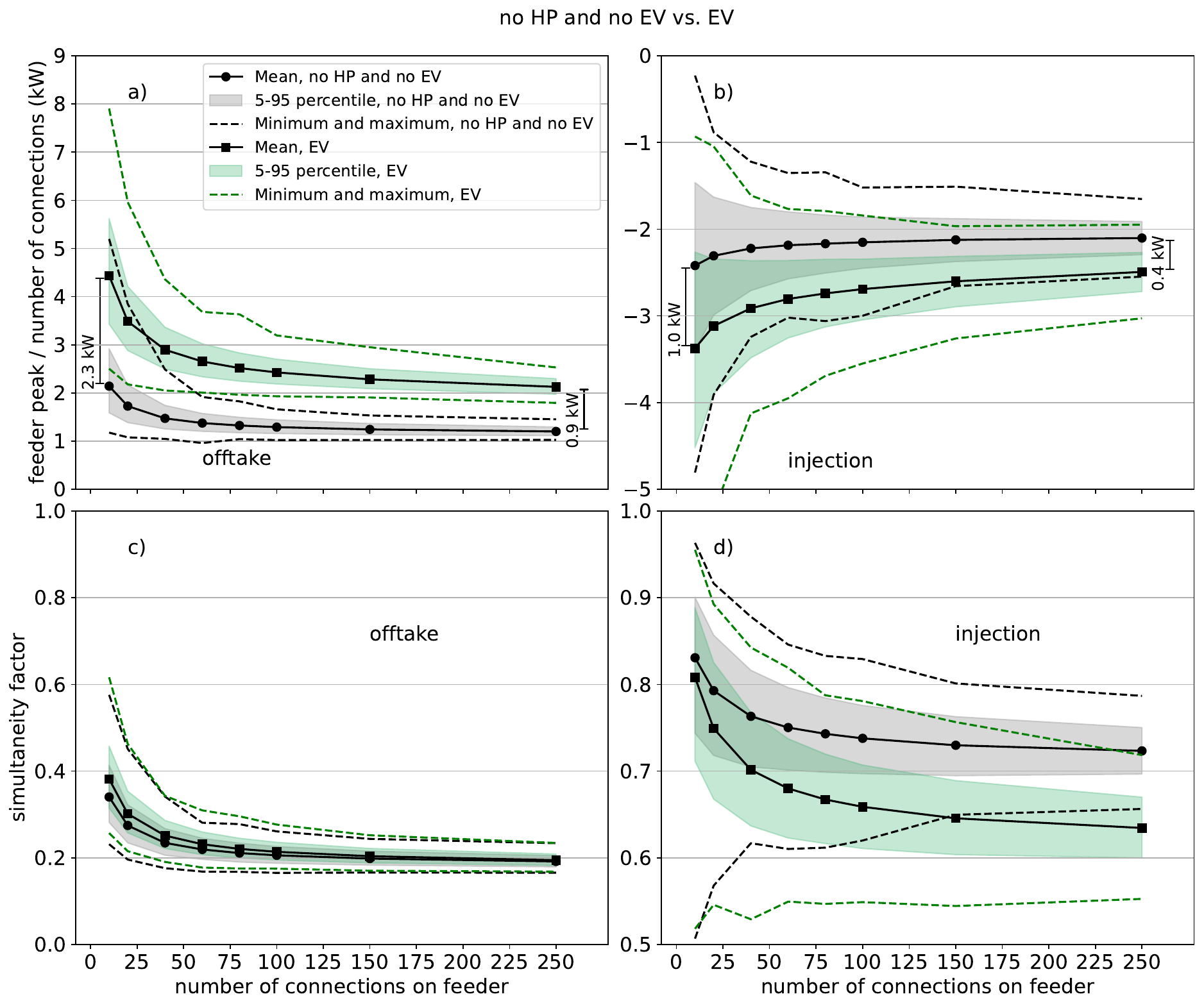}
\caption{Feeder peak per connection (kW)  (Eq.~\eqref{eq:feederpeak}) and simultaneity factor (Eq.~\eqref{eq:simfactor}) for \textit{no HP, no EV} feeders versus \textit{EV} feeders, for different number of connections, both offtake and injection.}
\label{fig:lct_diff_ev}
\end{figure}

A single EV adds between 2.3 kW (10 connections) and 0.9 kW (250 connections) to the feeder peak, as shown in Figure \ref{fig:lct_diff_ev}. The change in simultaneity factor, Figure \ref{fig:lct_diff_ev}c, is small to nonexistent. This is an illustration of the fact that the simultaneity factor by itself, which for DSOs is traditionally an important variable to estimate the necessary grid capacity, should be interpreted with care. Indeed, we observe an increase of the feeder peak that is caused solely by an increase in the magnitude of the individual peaks, without any change in simultaneity factor.

\begin{figure}
\centering
\includegraphics[width=1.\textwidth]{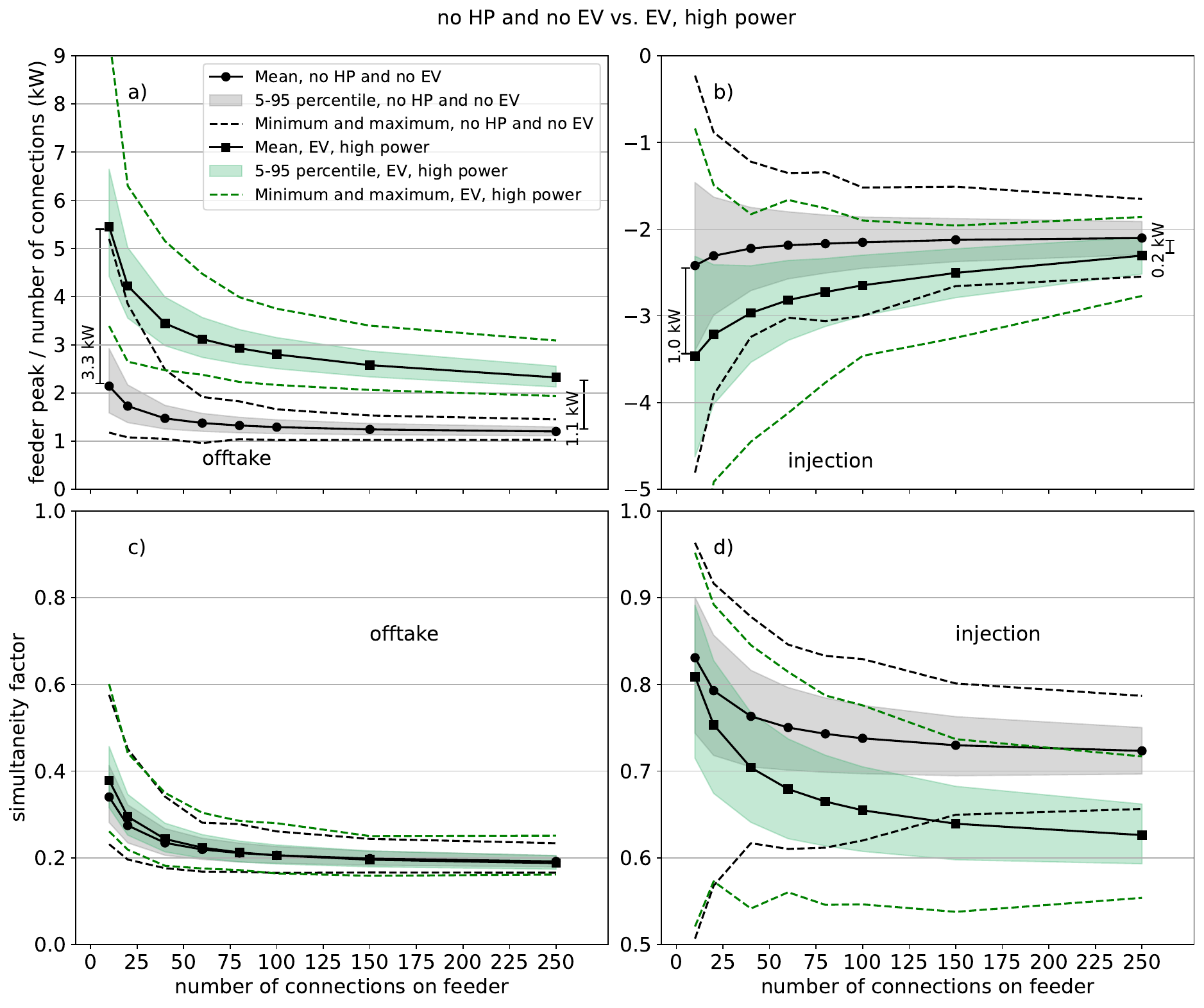}
\caption{Feeder peak per connection (kW)  (Eq.~\eqref{eq:feederpeak}) and simultaneity factor (Eq.~\eqref{eq:simfactor}) for \textit{no HP, no EV} feeders versus \textit{EV, high power} feeders, for different number of connections, both offtake and injection.}
\label{fig:lct_diff_evhigh}
\end{figure}

A single EV that charges at high power adds between 3.3 kW (10 connections) and 1.1 kW (250 connections) to the feeder peak, as shown in Figure \ref{fig:lct_diff_evhigh}. The change in simultaneity factor, Figure \ref{fig:lct_diff_evhigh}c, is small to nonexistent, similar to the result for EVs.

\subsection{Distribution of time of feeder peak}

We investigate at what time of the year the feeder peak occurs, and how this changes when LCTs are added to the LVG.

\subsubsection{Day of the year\label{sec:dayofyearpeak}}

\begin{figure}
\centering
\includegraphics[width=0.9\textwidth]{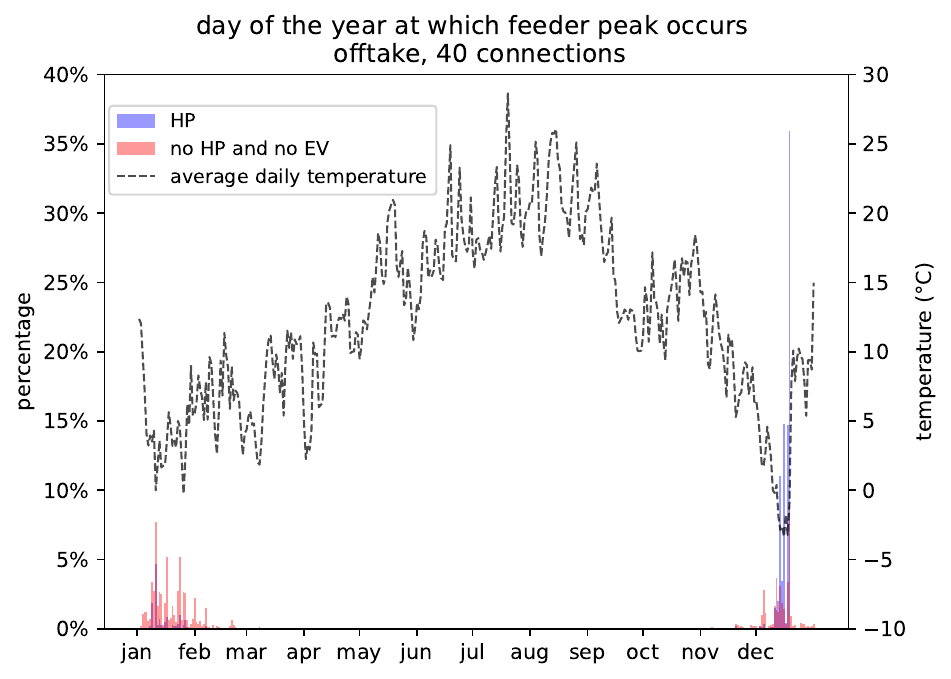}
\caption{Histogram of on what day of the year the feeder peak occurs, for \textit{no HP, no EV} and \textit{HP} feeders, for 40 connections, for offtake. The average daily temperature is plotted as well.}
\label{fig:doy_wp_nowp}
\end{figure}

In Figure \ref{fig:doy_wp_nowp} we plot the distribution of on what day the feeder peak occurs, for offtake. We compare \textit{no HP, no EV} with \textit{HP} feeders. For \textit{no HP, no EV} feeders, we see a large amount of variation on what day the peak occurs. It is always between November and March, and occurs mostly on colder days. For \textit{HP} feeders, peaks mostly occur on days with the lowest temperatures. Note that traditionally it is assumed that feeder peaks occur on the coldest day. Our results show that this assumption only holds when there is a large concentration of electric heating.

\begin{figure}
\centering
\includegraphics[width=0.9\textwidth]{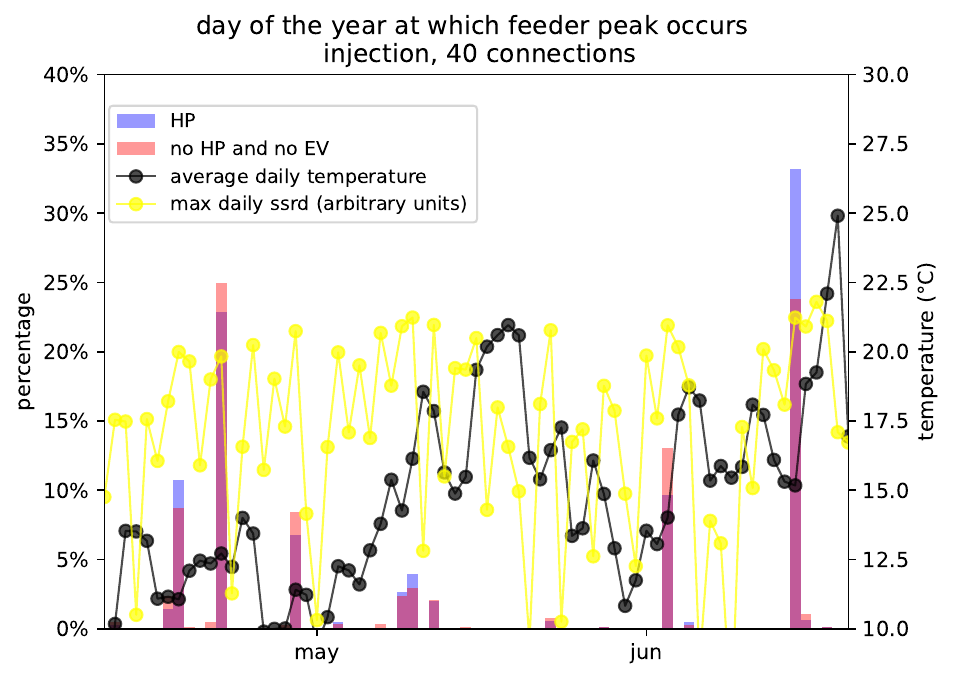}
\caption{Histogram of on what day of the year the injection feeder peak occurs for offtake, for HP and no HP and EV, for 40 connections. The average daily temperature and maximum surface solar radiation downwards (arbitrary units) is plotted as well. We have zoomed in on the months in which the feeder peaks occur}
\label{fig:doy_wp_nowp_inject_zoom}
\end{figure}

\begin{figure}
\centering
\includegraphics[width=0.9\textwidth]{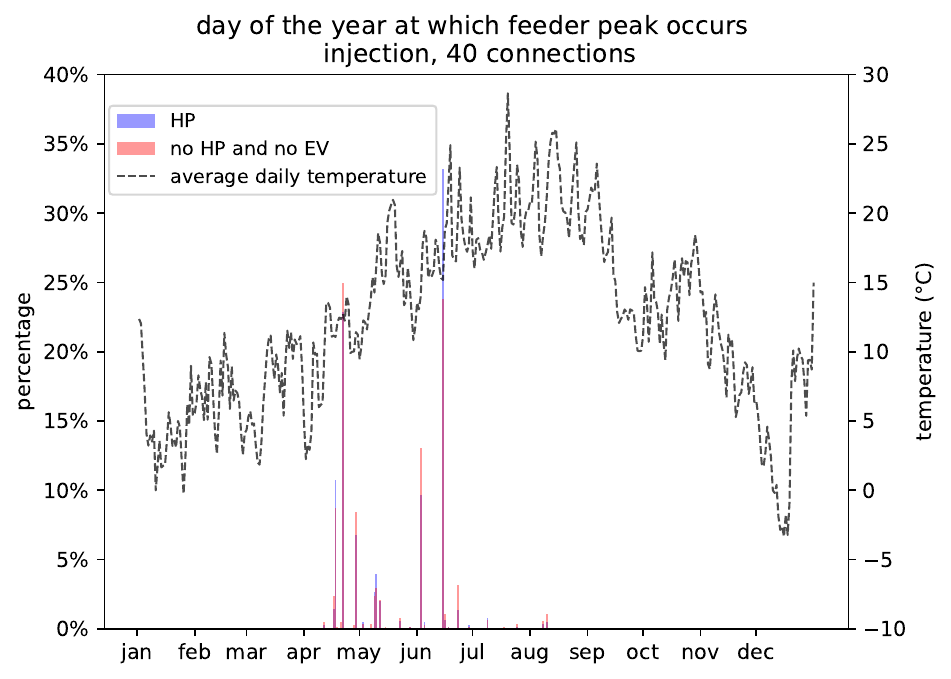}
\caption{Histogram of on what day of the year the injection feeder peak occurs, for \textit{no HP, no EV} and \textit{HP} feeders, for 40 connections. The average daily temperature is plotted as well.}
\label{fig:doy_wp_nowp_inject}
\end{figure}

A more detailed view is provided in Figure \ref{fig:doy_wp_nowp_zoom}. The day where the peak most often occurs is the final day of a cold spell. For \textit{EV} and \textit{EV, high power}, the distribution is similar to the one for \textit{no EV, no HP}, see Figure \ref{fig:doy_ev_evhigh} in \ref{app:doy}.
Injection peaks happen at similar days for all subsets, namely between April and August, see Figure \ref{fig:doy_wp_nowp_inject} for \textit{no HP, no EV} and \textit{HP} feeders. Injection peaks happen on days with a large maximal ssrd but when it is not too warm, as shown in Figure \ref{fig:doy_wp_nowp_inject_zoom}. Likely reasons are that the efficiency of solar panels decreases with increasing temperature, and that more air conditioning is running, which decreases the injection peak.
For \textit{EV} and \textit{EV, high power}, the distribution of when the injection peak occurs is similar (not shown). 

\subsubsection{Hour of the day}

\begin{figure}[H]
    \centering
    \begin{subfigure}[b]{0.495\textwidth}
        \includegraphics[width=\textwidth]{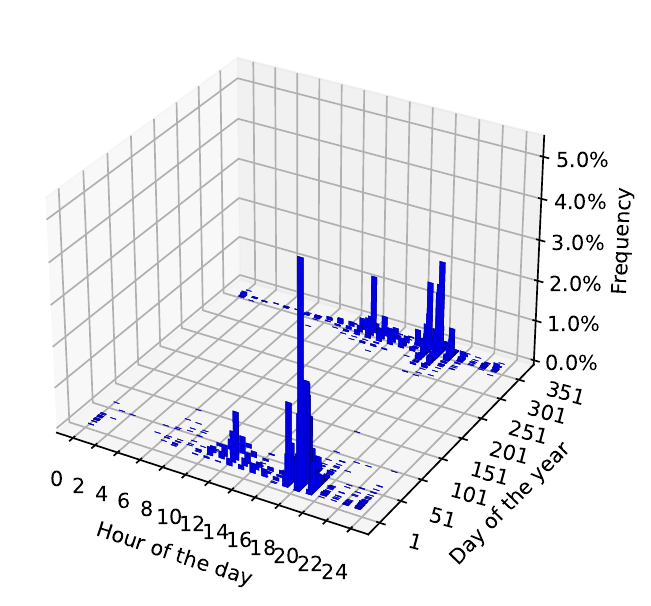}
        \caption{No HP, no EV}
        \label{fig:doy_off_sub1}
    \end{subfigure}
    \begin{subfigure}[b]{0.49\textwidth}
        \includegraphics[width=\textwidth]{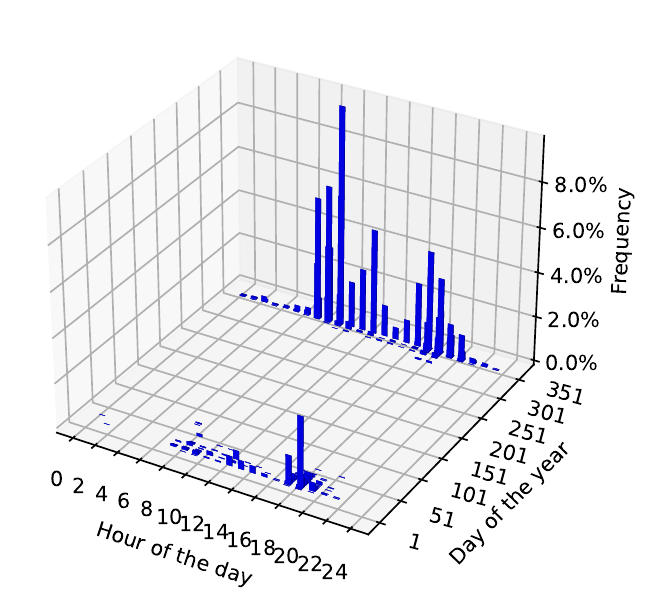}
        \caption{HP}
        \label{fig:doy_off_sub2}
    \end{subfigure}

    \begin{subfigure}[b]{0.495\textwidth}
        \includegraphics[width=\textwidth]{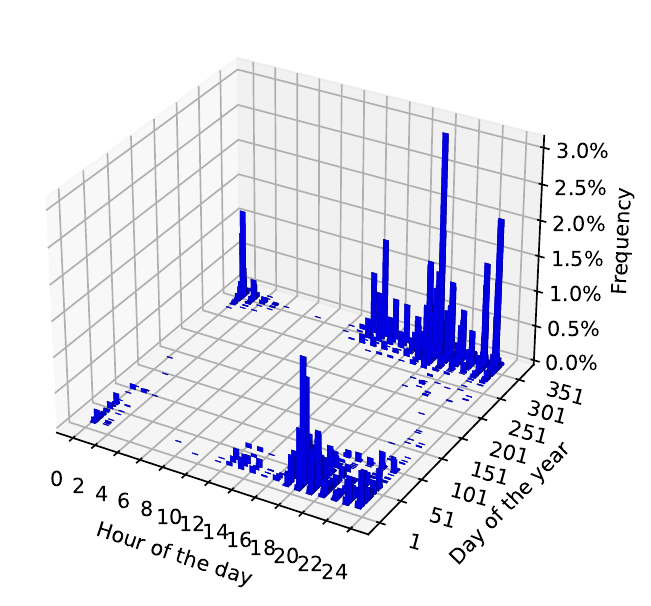}
        \caption{EV}
        \label{fig:doy_off_sub3}
    \end{subfigure}
    \begin{subfigure}[b]{0.49\textwidth}
        \includegraphics[width=\textwidth]{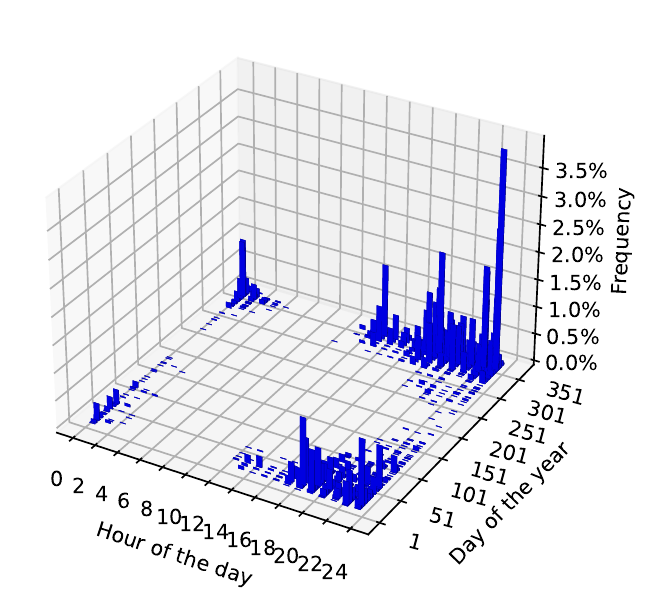}
        \caption{EV, high power}
        \label{fig:doy_off_sub4}
    \end{subfigure}
    \caption{Distribution of when the feeder peak occurs, for offtake, for 40 connections. The hour of the day is shown on the x-axis, and the day of the year is shown on the y-axis. The percentage of sampled feeders that had their peak at that moment is shown on the z-axis.}
    \label{fig:offtake_peakofyear}
\end{figure}

The distribution of the hour of the day and day of the year at which the feeder peak occurs is shown in Figure \ref{fig:offtake_peakofyear}, for the 4 different subsets. Results are for feeders with 40 connections, which are feeders with medium stochasticity, as discussed later. We see clear differences between the subsets. For \textit{no HP, no EV}, the peak typically occurs around 18h one a cold winter day. This is the time at which feeder peaks traditionally were assumed to occur, which is hereby confirmed by our results. Peaks also sometimes occur in the middle of the day in winter. 
For \textit{HP} feeders this changes to a morning (between 6h and 10h) and evening peak in winter. The most occurring peak timing is in the morning. Some midday peaks are also observed. The peaks are concentrated on cold days, which was already observed in Section \ref{sec:dayofyearpeak}.
For \textit{EV} feeders, peaks at around midnight (between 22h and 1h) in winter appear. Midday and evening peaks also occur, but no morning peaks are observed. Even in summer there is a small percentage of feeder peaks.
For \textit{EV, high power} feeders, the same timings are observed as for EVs. Interestingly, the most occurring peak is between 22h and 23h in winter. Peaks in the summer are still rare, but they occur more often compared to EVs, indicating that these are peaks caused by EV charging at high power, which are not temperature dependent.

\begin{figure}[H]
    \centering
    \begin{subfigure}[b]{0.495\textwidth}
        \includegraphics[width=\textwidth]{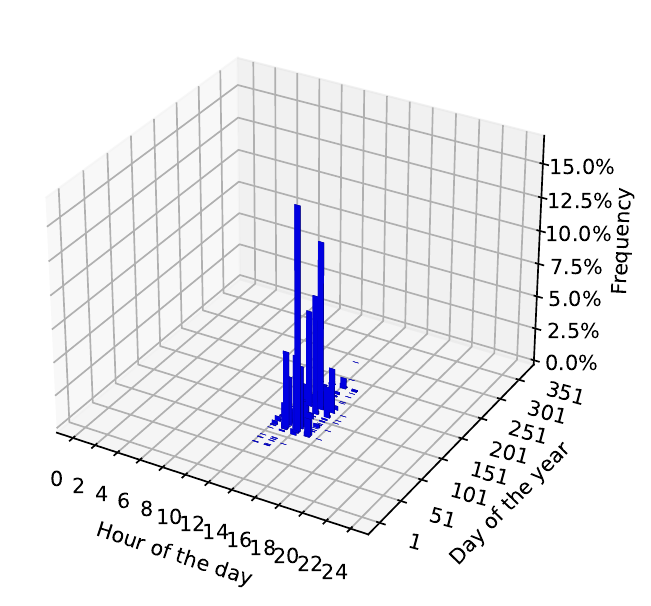}
        \caption{No HP, no EV}
        \label{fig:doy_inj_sub1}
    \end{subfigure}
    \hfill 
    \begin{subfigure}[b]{0.49\textwidth}
        \includegraphics[width=\textwidth]{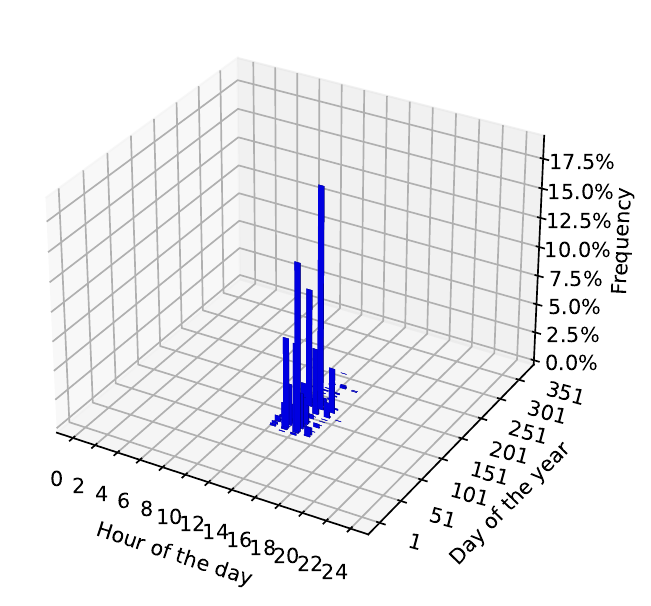}
        \caption{HP}
        \label{fig:doy_inj_sub2}
    \end{subfigure}

    \begin{subfigure}[b]{0.495\textwidth}
        \includegraphics[width=\textwidth]{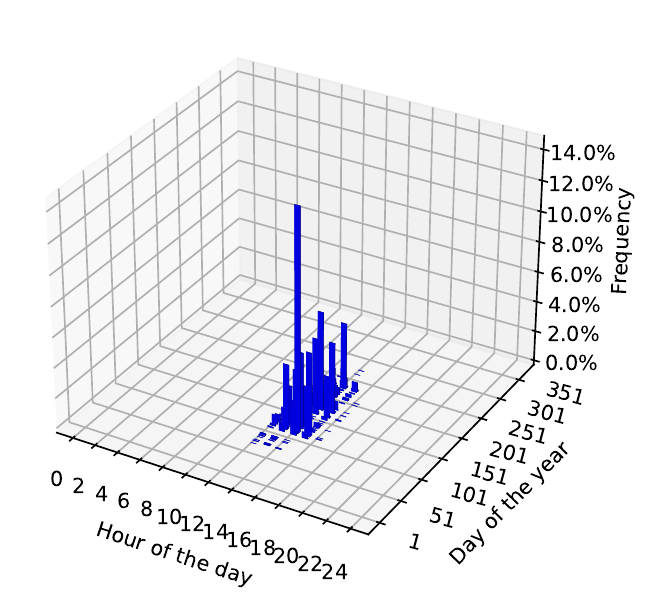}
        \caption{EV}
        \label{fig:doy_inj_sub3}
    \end{subfigure}
    \hfill
    \begin{subfigure}[b]{0.49\textwidth}
        \includegraphics[width=\textwidth]{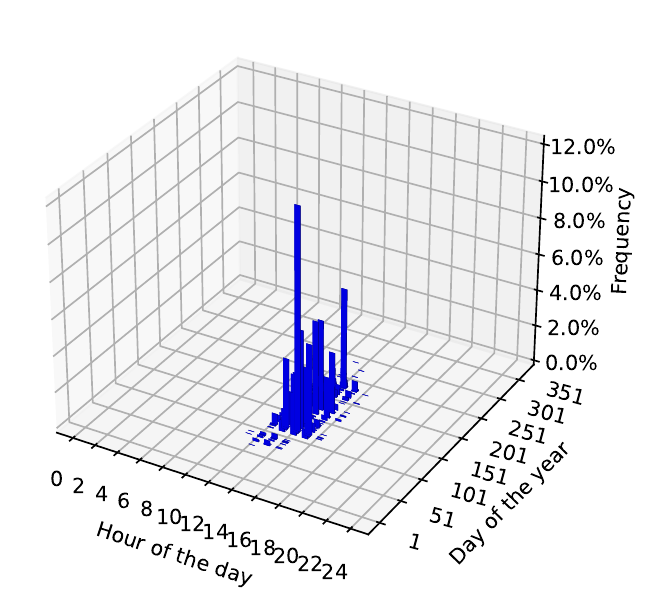}
        \caption{EV, high power}
        \label{fig:doy_inj_sub4}
    \end{subfigure}

    \caption{Distribution of when the feeder peak occurs, for injection, for 40 connections. The hour of the day is shown on the x-axis, and the day of the year is shown on the y-axis. The percentage of sampled feeders that had their peak at that moment is shown on the z-axis.}
    \label{fig:injection_peakofyear}
\end{figure}

The timing of the hour and day of year for injection of the 4 different subsets is shown in Figure \ref{fig:injection_peakofyear}. All subsets show similar behavior: peaks occur mostly between 13h and 16h in the spring and summer months (April to September).

\begin{figure}[H]
    \centering
    \begin{subfigure}[b]{0.495\textwidth}
        \includegraphics[width=\textwidth]{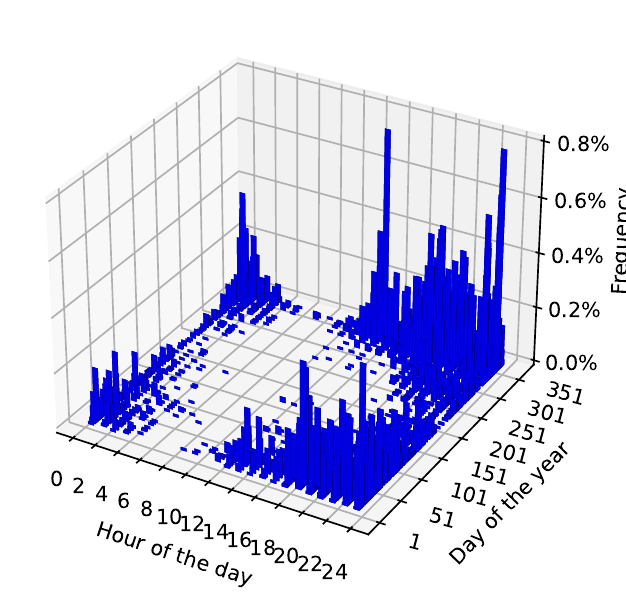}
        \caption{EV, high power, 10 connections}
        \label{fig:doy_inj_sub1_nconn}
    \end{subfigure}
    \begin{subfigure}[b]{0.49\textwidth}
        \includegraphics[width=\textwidth]{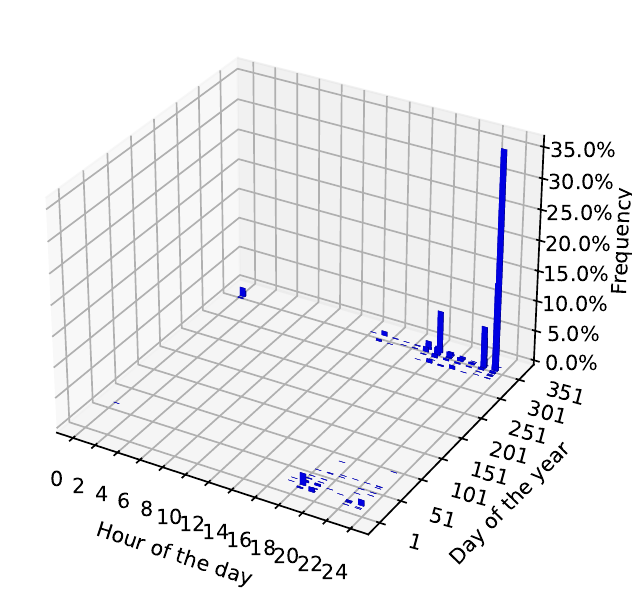}
        \caption{EV, high power, 250 connections}
        \label{fig:doy_inj_sub2_nconn}
    \end{subfigure}

    \begin{subfigure}[b]{0.495\textwidth}
        \includegraphics[width=\textwidth]{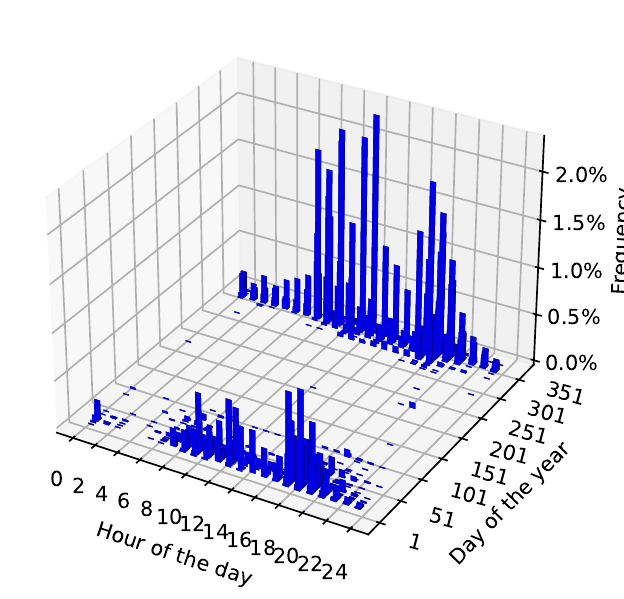}
        \caption{HP, 10 connections}
        \label{fig:doy_inj_sub3_nconn}
    \end{subfigure}
    \begin{subfigure}[b]{0.49\textwidth}
        \includegraphics[width=\textwidth]{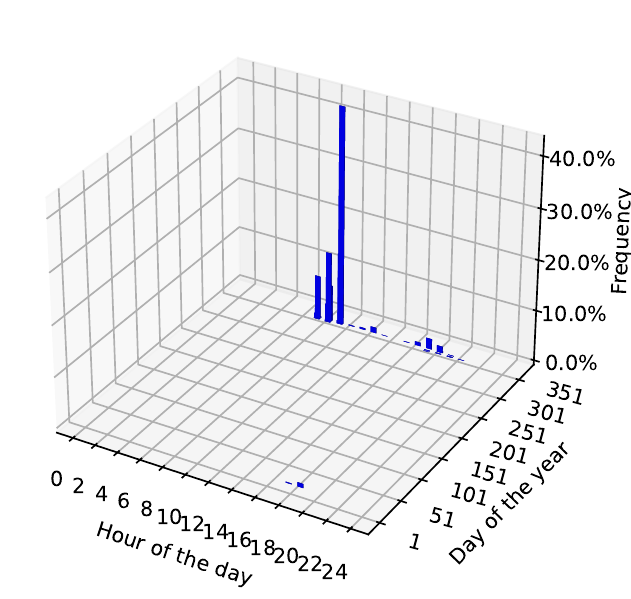}
        \caption{HP, 250 connections}
        \label{fig:doy_inj_sub4_nconn}
    \end{subfigure}
    \caption{Distribution of when the feeder peak occurs, for offtake, for \textit{HP} and \textit{EV, high power} feeders, for 10 and 250 connections. The hour of the day is shown on the x-axis, and the day of the year is shown on the y-axis. The percentage of sampled feeders that had their peak at that moment is shown on the z-axis.}
    \label{fig:offtake_nconn}
\end{figure}

The influence of feeder size on the timing of the feeder peaks is illustrated in Figure \ref{fig:offtake_nconn}. The distribution for feeders with 10 and 250 connections is compared for \textit{EV, high power} feeders and \textit{HP} feeders. For the EVs, feeder peaks can occur on any day of the year for feeders with 10 connections. For 250 connections, the distribution is almost completely concentrated on one moment (22h on the last day of a cold spell). This illustrates that small feeders are highly stochastic in their behavior. This effect is the most pronounced for EVs.
For HP feeders, there is less variation in the days on which the feeder peak occurs for 10 connections, where it is still strongly determined by low temperatures. For 250 connections, almost all feeder peaks happen on the same day, which is the last day of a cold spell.

\subsection{Relation of feeder consumption to weather\label{sec:feederplots}}

We investigated the behavior of feeder loads throughout the day, and their qualitative dependence on the weather. We plotted the feeder consumption as a function of time, together with the temperature, amount of sunshine, and probability to see the feeder peak that day of the year. The distribution of the power at each quarter hour was calculated from the same 10,000 samples used to generate the other results, as discussed in Section \ref{sec:methods_sampling}. Arrows point at the time of day at which the feeder peak most often occurs \textit{on that day}.

\begin{figure}
\includegraphics[width=\textwidth]{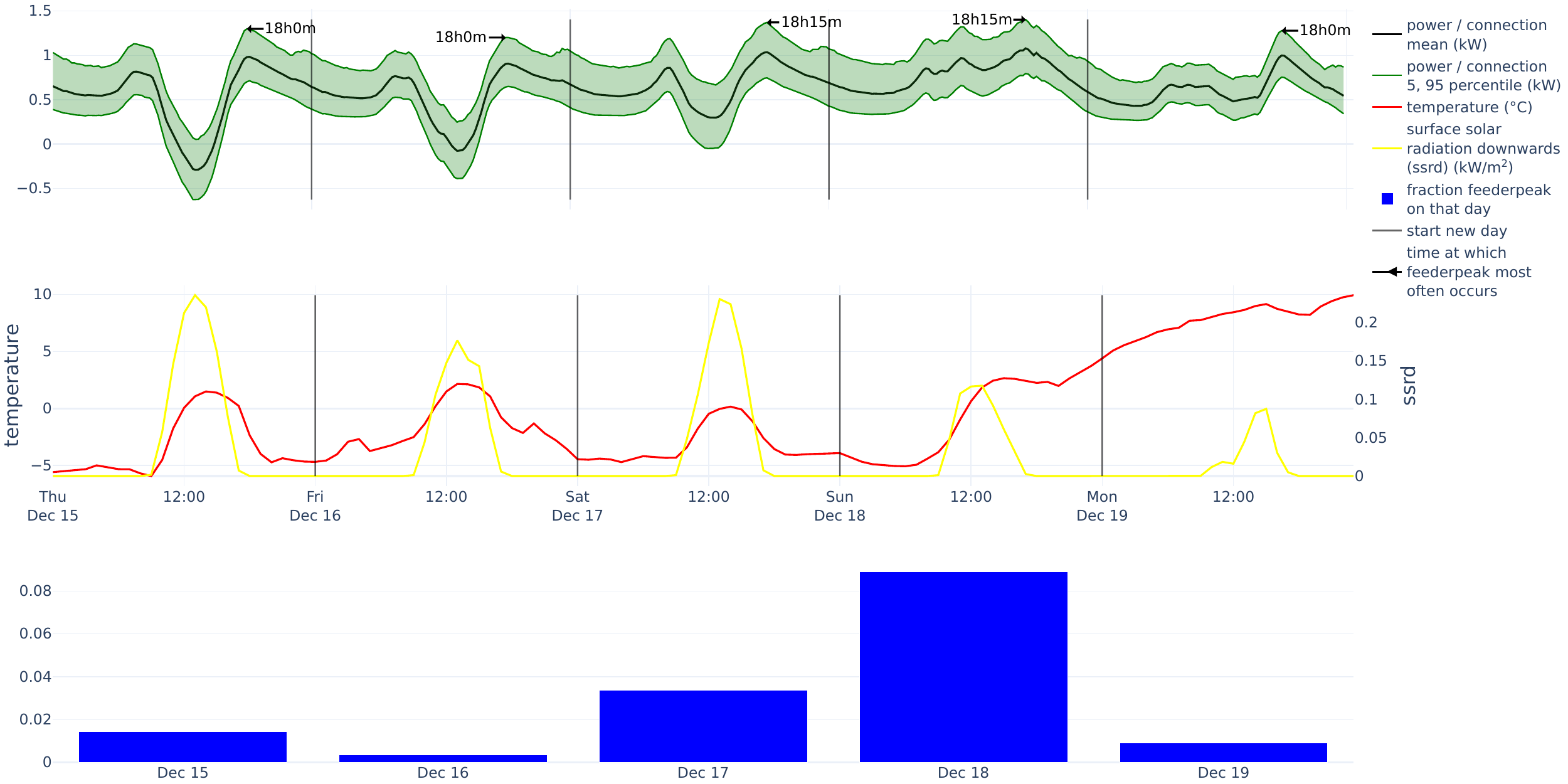}
\caption{Histogram of consumption on feeder level, together with temperature and solar irradiation and probability to have the highest feeder peak that day. For \textit{no EV, no HP}.}
\label{fig:fw_no_hp_no_ev}
\end{figure}

In Figure \ref{fig:fw_no_hp_no_ev} we plot the coldest day and surrounding days of 2022, for the \textit{no HP, no EV} subset. We see a classical duck curve shape \cite{denholm2015overgeneration}, with the daily peak most often occurring around 18h00. There is less sunshine on the $18^{th}$ and $19^{th}$ December, leading to a curve that is much flatter compared to the duck-curves on previous days. PV has a significant influence, even on these cold winter days, as in Belgium the coldest days are typically also unclouded sunny days. Net consumption can even be negative, such as on December $15^{th}$ during the middle of the day. The day where the feeder peak happens most often is the last day of a cold spell. It was a very cold morning and there wasn't a significant amount of PV production.

\begin{figure}
\includegraphics[width=\textwidth]{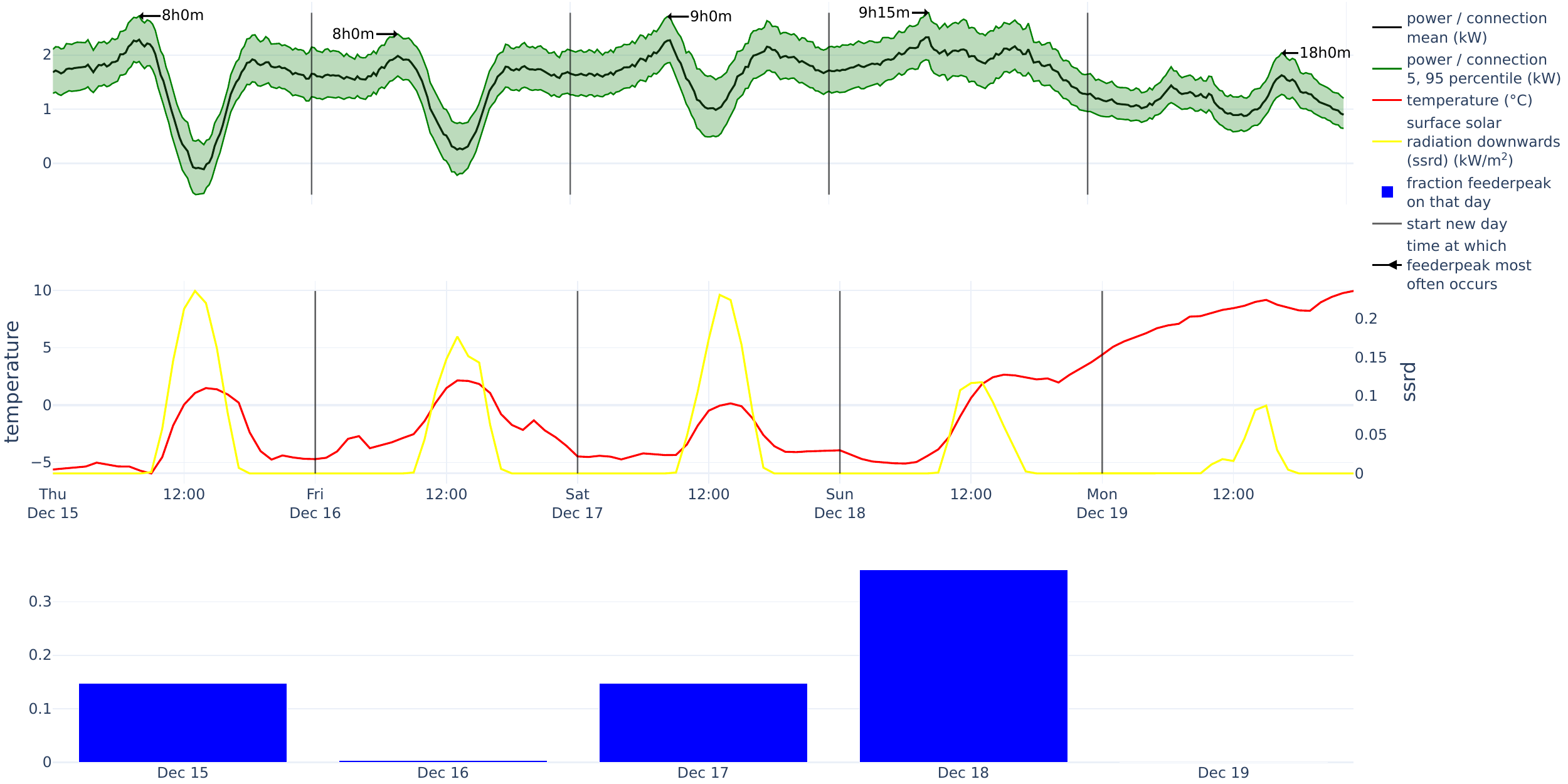}
\caption{Histogram of consumption on feeder level, together with temperature and solar irradiation and probability to have the highest feeder peak that day. For HP.}
\label{fig:fw_hp}
\end{figure}

For heat pumps, Figure \ref{fig:fw_hp}, the peaks in the morning and evening are of similar magnitude, with the mornings being slightly higher than the evenings. PV production has a large influence: consumption during the middle of the day is small on days with a large amount of sunshine. Injection is lower compared to the \textit{no HP, no EV} feeders, likely because of increased consumption of the HPs during the day. The feeders behave like a ``night-camel'', with two humps during the morning and evening peak, and a rather high back during the night. On the day on which the feeder peak most often occurs, the feeder consumption profile is almost completely flat. It is the final day of a cold spell with minimal PV production (the same day as for the \textit{no HP, no EV} subset).

\begin{figure}
\includegraphics[width=\textwidth]{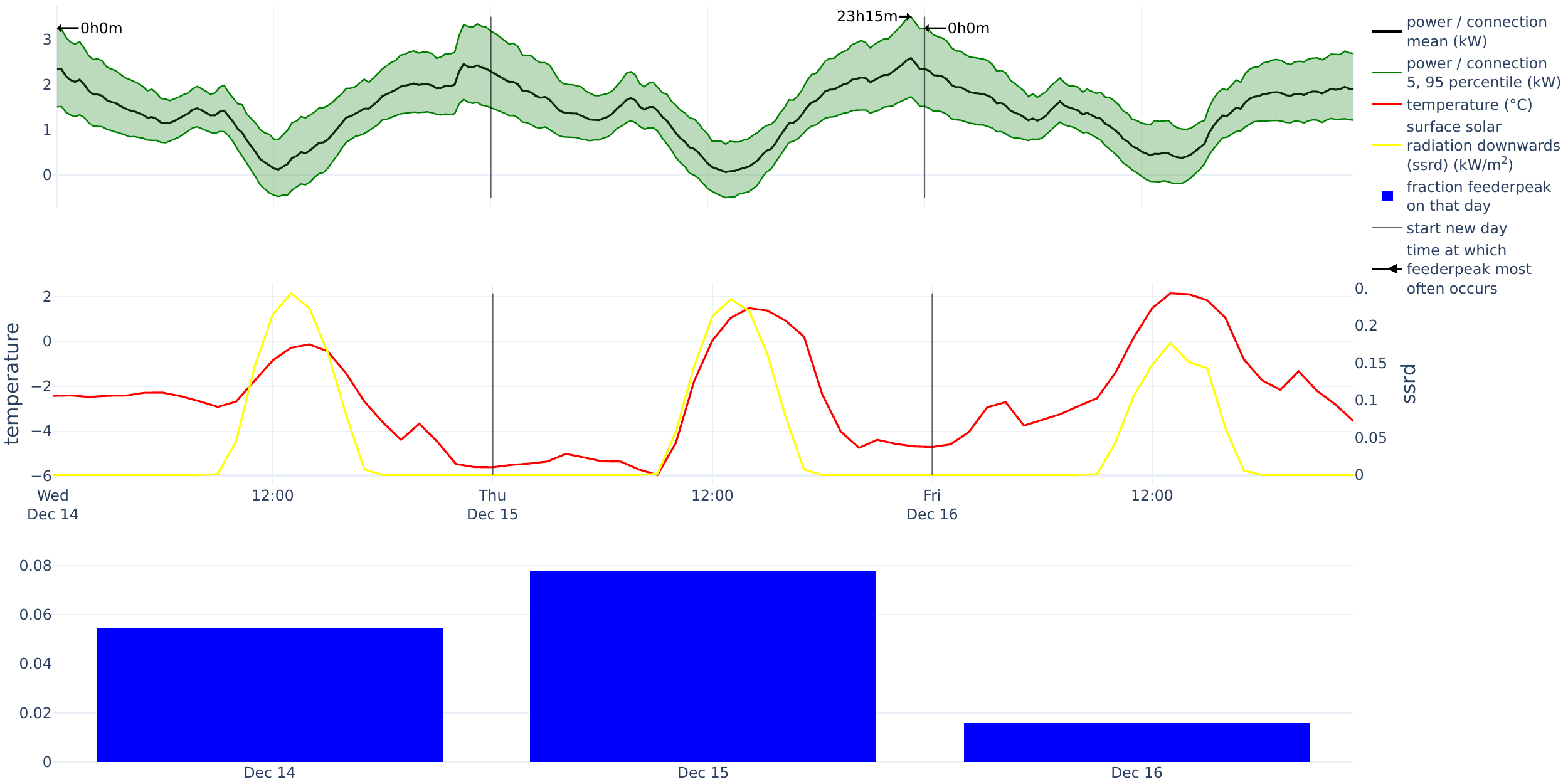}
\caption{Histogram of consumption on feeder level, together with temperature and solar irradiation and probability to have the highest feeder peak that day. For EV high power.}
\label{fig:fw_evhigh}
\end{figure}

For the EVs with high charging power, Figure \ref{fig:fw_evhigh}, we see a ``night-dromedary'' curve, with a single hump around midnight and the head of the dromedary as a smaller morning peak. Most daily feeder peaks occur between 22h and 1h, with a small local peak in the morning. 
Our hypothesis is that this shape is caused by the strongly ingrained habits in Belgium to shift consumption to the moments when there is a lower night tariff. Charging behavior in other countries can be different. Also, as the cost reduction of the night tariff has recently strongly decreased, it is possible that this behavior still evolves in the years ahead.

\begin{figure}
\includegraphics[width=\textwidth]{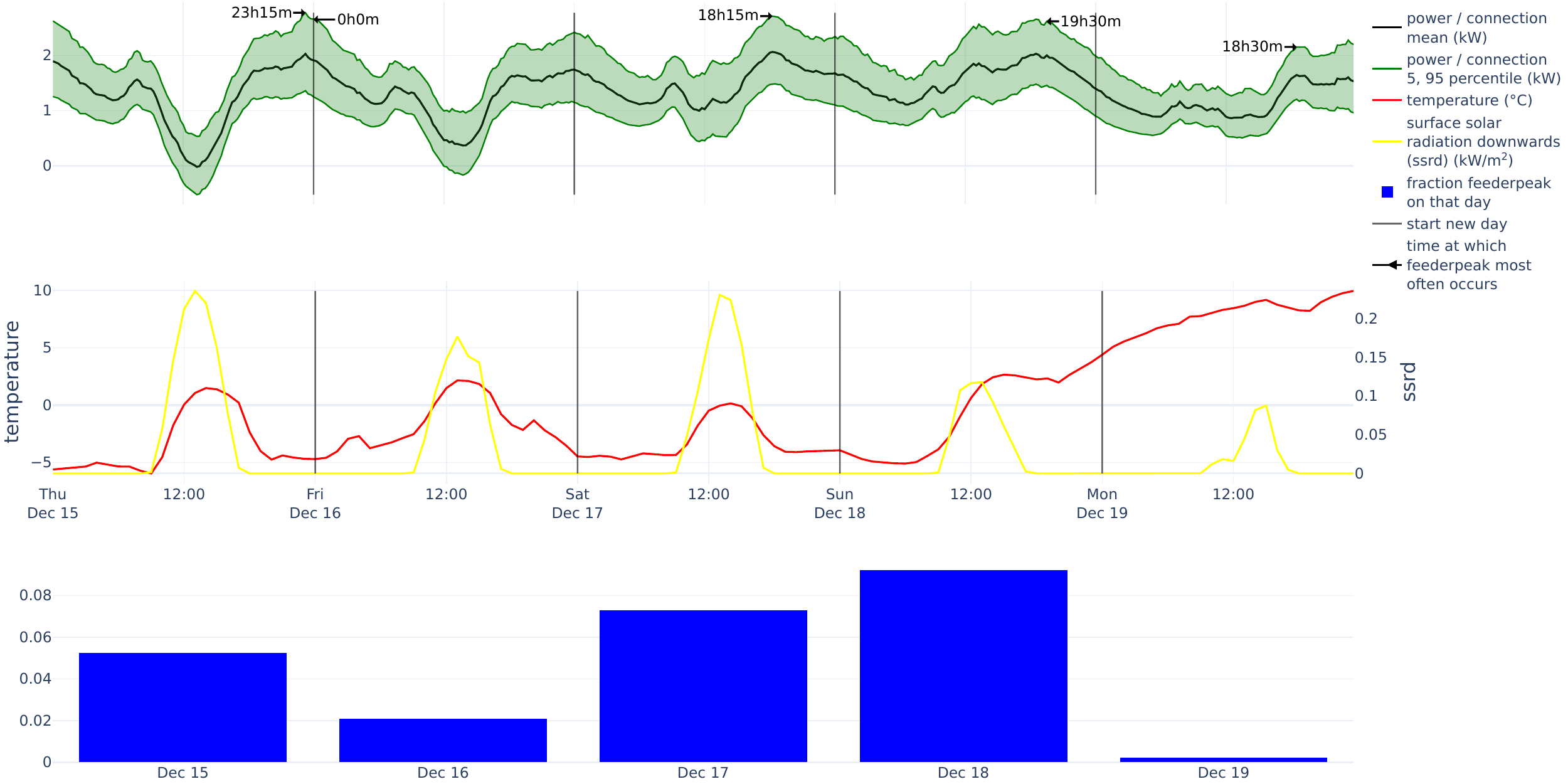}
\caption{Histogram of consumption on feeder level, together with temperature and solar irradiation and probability to have the highest feeder peak that day. For EV.}
\label{fig:fw_ev}
\end{figure}

For the EVs, the situation is somewhere between \textit{no HP, no EV} and \textit{EV, high power}, with a mix of duck and night-dromedary behavior, see Figure \ref{fig:fw_ev}.

\begin{figure}
\includegraphics[width=\textwidth]{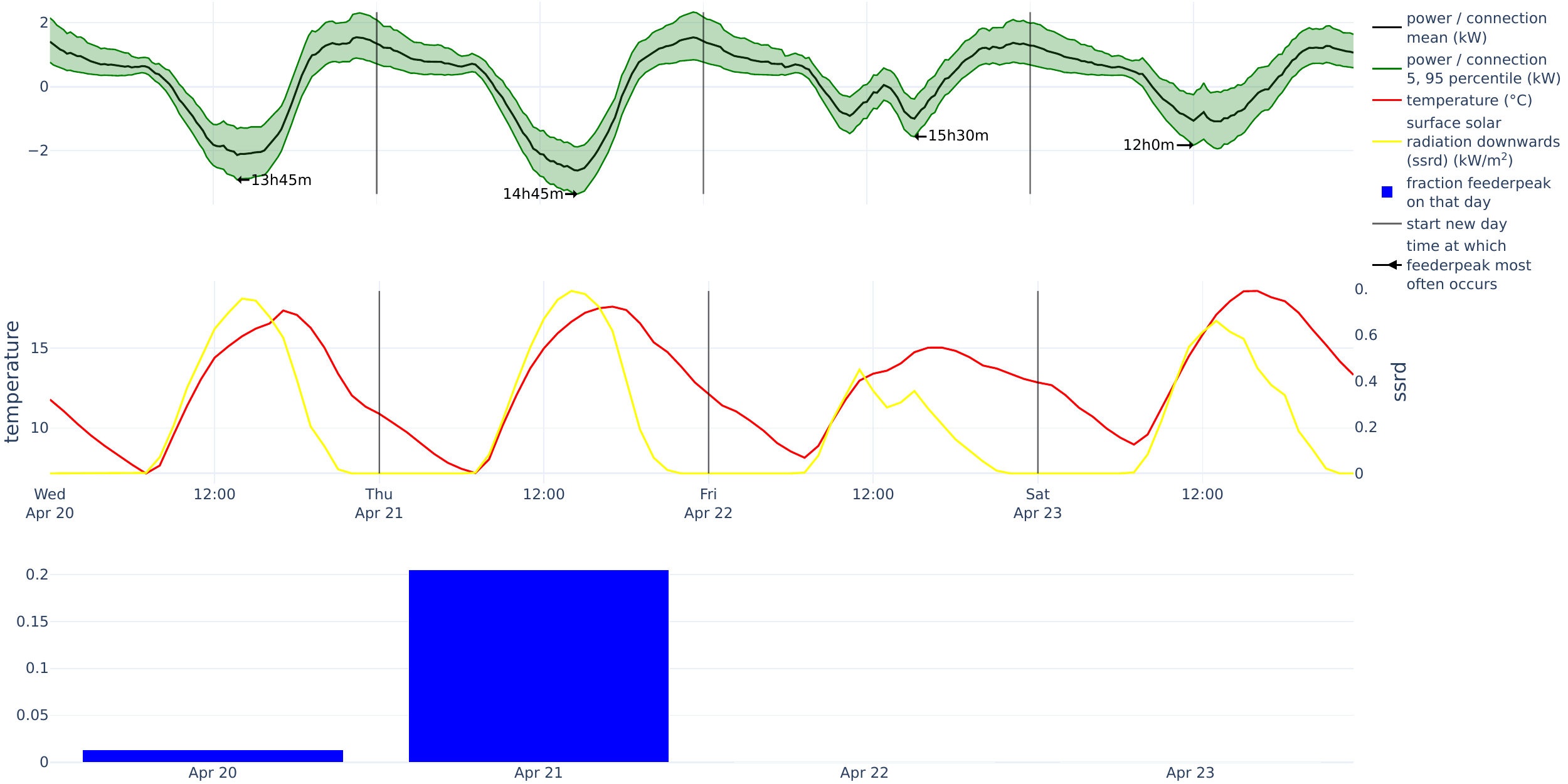}
\caption{Histogram of consumption on feeder level, together with temperature and solar irradiation and probability to have the highest feeder injection peak that day. For HP.}
\label{fig:fw_hp_summer}
\end{figure}

In Figure \ref{fig:fw_hp_summer} we plot the HP feeder curve for a moment of peak injection. We observe close-to-zero consumption, with large injection during the day. Injection peaks typically occur between 13h and 15h. With injection peaks per connection of around 4 kW, these houses deliver a significant amount of energy to the grid.

\section{Discussion\label{sec:discussion}}

\subsection{Feeder peak and simultaneity factor}

The contribution to the feeder peak of HPs and EVs is significant, as expected. For typical feeder sizes (10 to 40 connections), HPs add less to the feeder peak than EVs, and EVs add less than EVs that charge at higher power, which is also as expected. A surprising result is that the effect on the feeder peak of a HP or an EV is similar for 250 connections, which is at the medium-voltage to low-voltage transformer level. For typical feeder sizes (10 to 40 connections), simultaneity factors were similar among the subsets, with the \textit{no HP, no EV} subset (approximately 0.35) being lower than the other subsets (approximately 0.4). At 250 connections, HP feeders have a simultaneity factor of approximately 0.25, while the other three subsets have a value of around 0.2. This indicates that HP consumption is more correlated among different connections than EV charging. Since HP consumption is strongly correlated with cold weather, this result is understandable. The low correlation (i.e., low synchronization) of EV charging was less expected. For typical feeder sizes (10 to 40 connections), simultaneity factors for injection are between 0.8 and 0.85 for the subsets. At 250 connections, HP feeders have a simultaneity factor for injection of around 0.75, while for \textit{EV} and \textit{EV, high power} this value is around 0.65. This significantly lower simultaneity factor is likely caused by EV charging during the day, which is done to increase self-consumption of PV production. There are large differences in injection feeder peaks. These mostly depend on the amount and size of installed PV, which differs for the subsets, see Table \ref{tab:dataset}.

It is important to note that these consumption profiles were measured before the introduction of a capacity tariff in the Flanders region on January 2023. The capacity tariff ties the distribution costs for residential customers to their average monthly peak consumption. It is expected that this incentive will lower the individual customer peaks. 

Injection feeder peaks caused by PV have a similar magnitude than offtake peaks. Offtake peaks are higher on average than injection peaks for feeders with EVs, but lower for feeders without EVs. The distribution of the injection and offtake feeder peaks show significant overlap. We conclude that it is not clear whether in the future the majority of LV feeders will be constrained by offtake or injection. This depends on the relative installment speed and sizing of PV installations versus HPs and EVs. Consumption behavior such as EV charging and HP usage also has an influence. Consumption behavior is influenced by policy, such as (possibly time-dependent) energy- and capacity tariffs.

Simultaneity of PV is typically expected to be close to 1. We find average values between 0.85 (10 connections) and 0.65 (250 connections), which is significantly lower than 1. The modeling approach is partly responsible for this. We draw profiles from the same day for the whole region of Flanders, but the amount of sunshine is locally more correlated. Some regions will have more sunshine while other places have less sunshine. These profiles are now sampled together, lowering the simultaneity factor for injection. This effect also causes and underestimation of the injection peak. It is not clear how large this effect is, although we expect it to be minor, as yearly injection peaks often happen on days when it is sunny in the whole of Flanders.

The morning peak caused by HPs was also found by Protopapadaki et al.~\cite{PROTOPAPADAKI2017268} and Love et al.~\cite{LOVE2017}. Our results provide further confirmation of their work.

\subsection{Possible behavior recommendations}

The observed influence of LCTs to the shifting of the timing of the feeder peak can inform recommendations on electricity consumption behavior.

Very cold days typically have clear skies with a significant amount of PV production, as seen in the Figures in Section \ref{sec:feederplots}. Strain on the grid could be alleviated if EVs are charged during the day on cold days.

Feeders with a large amount of heat pumps have a feeder peak in the morning, as seen in Figure \ref{fig:fw_hp}. If households don't let the temperature of their house drop too much at night, or if they shift sanitary hot water heating with the HP away from the morning and evening peak, this would reduce the feeder peaks.

As injection peaks play a significant role, households could shift consumption in summer to the middle of the day, to reduce their injection peaks.

Shifting EV charging to night times could become problematic, as we already see in our current data that offtake peaks happen at night, as shown in Figure \ref{fig:doy_off_sub4}.

\subsection{Dataset representativeness\label{sec:datarep}}

Our dataset is not representative of the general population in Flanders. The set of connections with smart meter profiles at the quarter-hour level consists of connections that have decided themselves to activate the measurement of their quarter-hour data. These are biased towards houses that are more recently build, have more installed LCTs, and have inhabitants with a stronger interest in monitoring and changing their electricity consumption behavior. This bias is mostly to our advantage, as we are interested in investigating future scenarios where most houses are fully electrified. These will typically have PV and decent to good insulation. Nevertheless, for the most likely future scenarios, we expect our dataset to contain too much hybrid EVs compared to full EVs. How actual EV charging behavior will evolve remains to be seen, and will likely show variations depending on the region. For example, dynamic tariffs could lead to synchronized charging at the lowest price point. Furthermore, heat pumps will be installed in houses that are, on average, less insulated compared to our dataset. These will contain less floor heating as well. This will cause an underestimation of the contribution to the feeder peak of HPs. Another source of underestimation of HP peaks is that we expect colder days to occur than what has been observed in 2022.

More generally, although these results provide data-driven insights as to how behavior is evolving, it is important to not blindly extrapolate the results from this dataset from 2022 towards 2030 or 2050, nor for other regions in Europe. The technical capacity to analyze the residential electricity consumption and production patterns, with methods as presented in this paper, is an important technical capability for modern European DSOs. It can be used to keep track of changing behavior on the LVG.

\subsection{Modeling assumptions\label{sec:modelassump}}

\begin{figure}
\includegraphics[width=\textwidth]{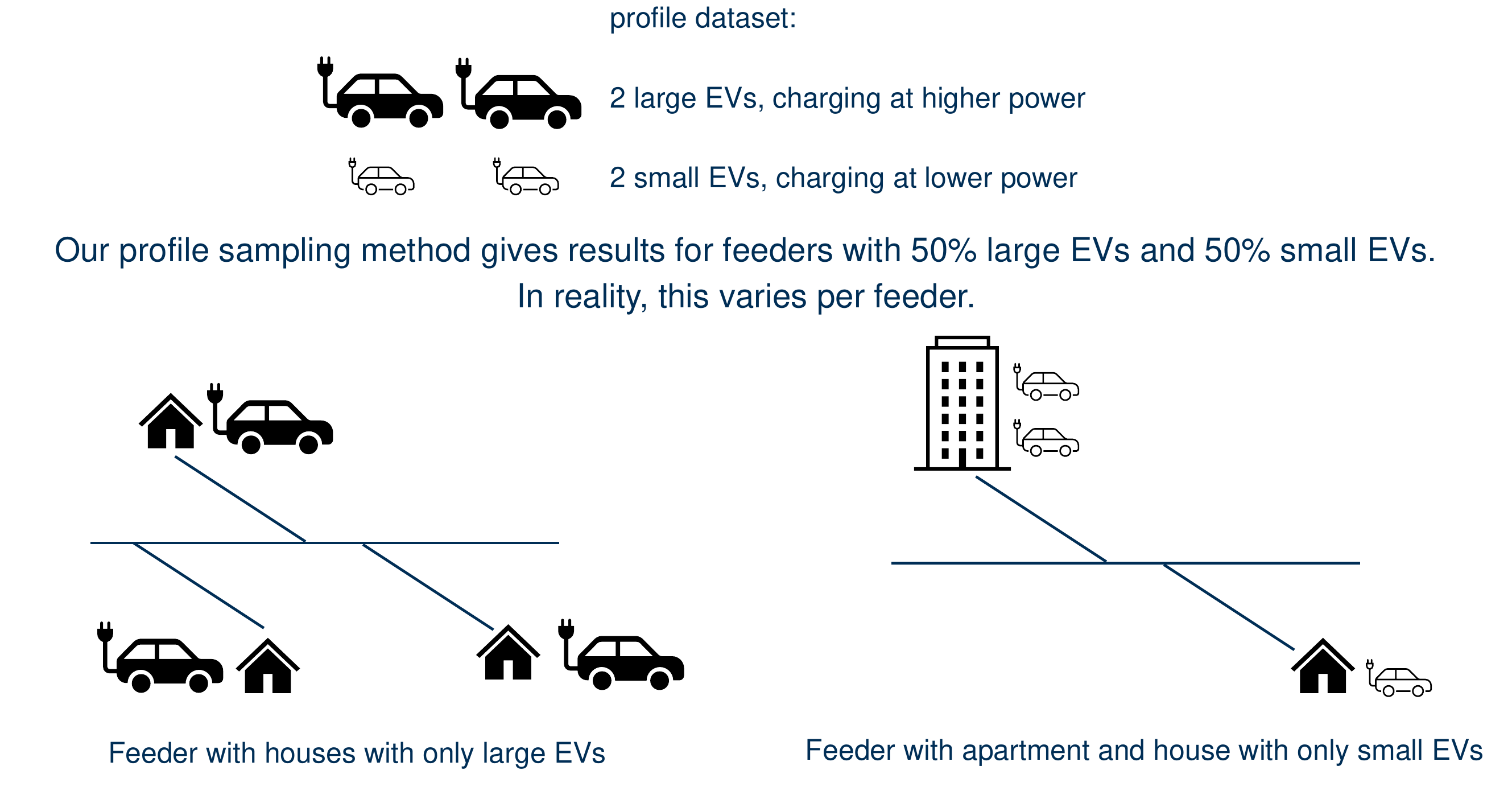}
\caption{Our modeling approach assumes that each feeder has the same distribution of consumption patterns as the profile data. This is not true, as illustrated in this figure. While the profile dataset has a certain percentage of EVs that will charge at higher power or lower power, we expect significant variations between feeders. For example a feeder with only large EVs (bottom left), or a feeder with an apartment and house containing only small EVs (bottom right).}
\label{fig:feeder_vs_profile}
\end{figure}

By sampling the profiles randomly, we assume that the distribution of connections on the feeder level is the same as the distribution of the available smart meter profiles. This is not true. For example, rich neighborhoods might have EVs with higher-power EV chargers and better insulated houses. A better estimate of the peak contribution of EVs in specific neighborhoods could be obtained by a more directed sampling, compared to our random sampling. This assumption is visualized in Figure \ref{fig:feeder_vs_profile}. Correcting for this assumption implies a more involved modeling, since a more directed sampling needs to be done for each connection, see e.g.~\cite{SOENEN2023100985}.

\begin{figure}
\includegraphics[width=\textwidth]{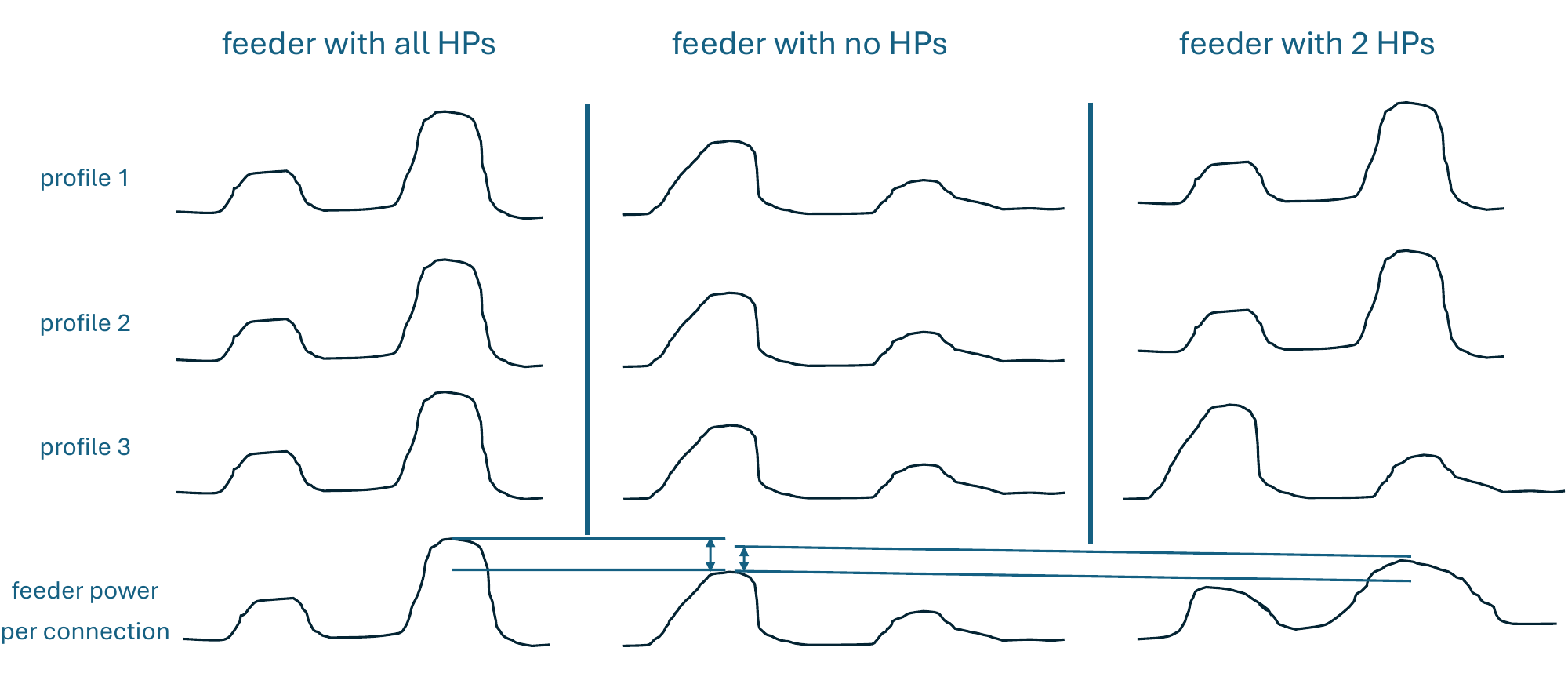}
\caption{When estimating the contribution of an HP (EV) to the feeder peak, as in Figures \ref{fig:hp_add}, \ref{fig:lct_diff_ev} and \ref{fig:lct_diff_evhigh}, we compare the difference in feeder peak per connection for feeders containing either no HP (EV) or all HP (EV). If not all connections have an HP (EV), this is only an approximation, since feeder peaks with or without an HP (EV) happen at different times. This is illustrated in this figure for a feeder of 3 connections, with all HPs, no HPs, and two HPs.}
\label{fig:not_all_hp}
\end{figure}

When estimating the contribution of an HP (EV) to the feeder peak, as in Figures \ref{fig:hp_add}, \ref{fig:lct_diff_ev} and \ref{fig:lct_diff_evhigh}, our estimates are an overestimation if not all connections have an HP (EV), since the peak of connections with and without an HP (EV) happen at a different time. This is illustrated in Figure \ref{fig:not_all_hp}.

\begin{figure}
\centering
\includegraphics[width=0.6\textwidth]{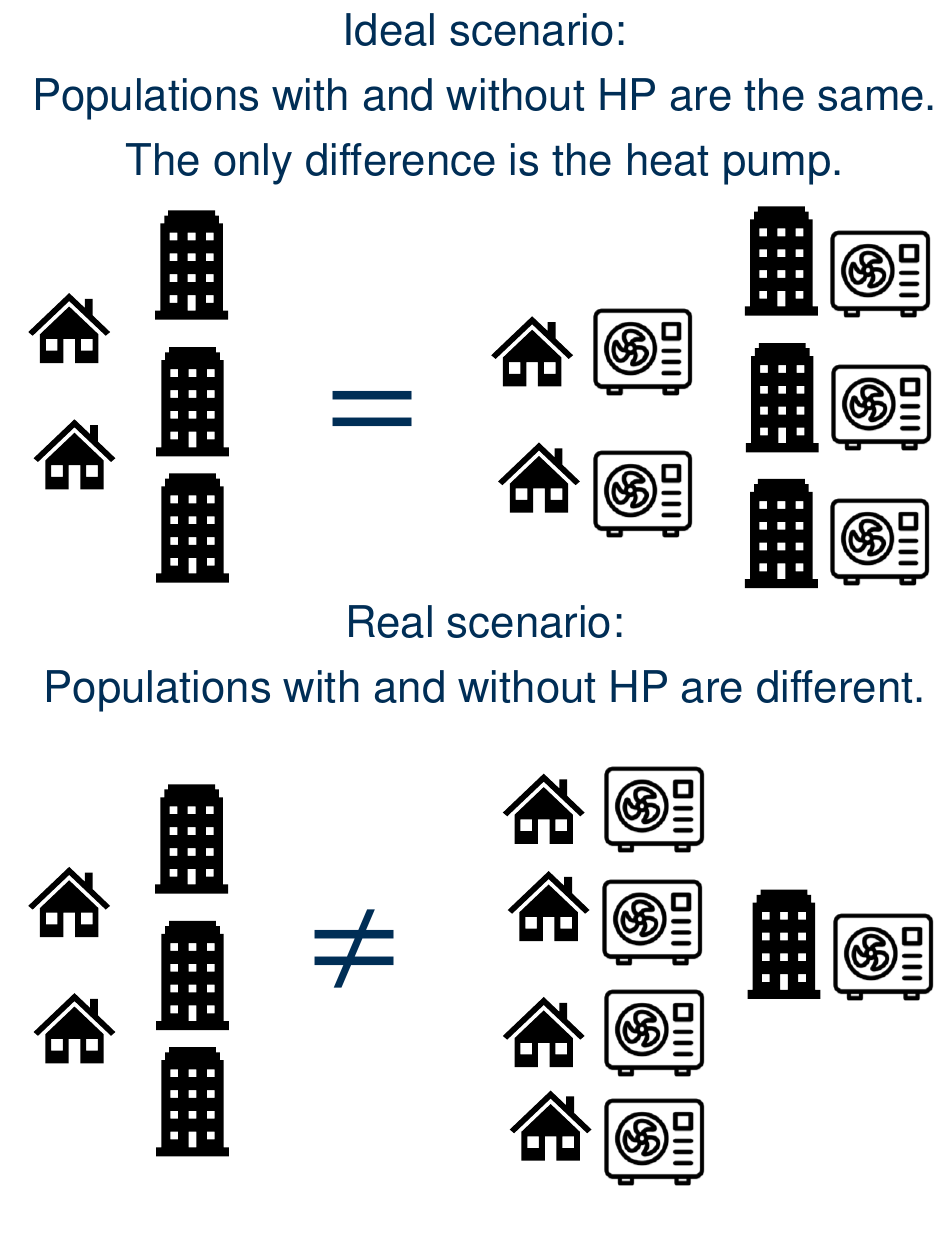}
\caption{To estimate the contribution of EVs and HPs to the feeder peak, as in Figures \ref{fig:hp_add}, \ref{fig:lct_diff_ev} and \ref{fig:lct_diff_evhigh}, we assume the populations with or without a HP (EV) are the same. This is only an approximation. In this figure, this is illustrated for populations with and without a HP, where the two categories in the population are residential houses and apartments.}
\label{fig:diff_pop}
\end{figure}

To estimate the contribution of EVs and HPs to the feeder peak, as in Figures \ref{fig:hp_add}, \ref{fig:lct_diff_ev} and \ref{fig:lct_diff_evhigh}, we compare the results from two different populations. Our method assumes that the population of connections with a HP (EV) is similar to the population of connections without a HP (EV), where the only difference is that a HP (EV) is installed, as illustrated in Figure \ref{fig:diff_pop}. This is likely not the case.
Given the limited access we have to information about the houses, correction for differences in the population distribution is not feasible. Furthermore, privacy considerations would likely make it infeasible to acquire the relevant variables needed to perform such a correction.

These modeling assumptions should be kept in mind when interpreting the results. We do not expect them to materially influence our qualitative results. The quantitative results on feeder peaks and simultaneity factors could be influenced to a significant degree.

\section{Conclusions\label{sec:conclusion}}

We have presented a simple yet insightful profile sampling method to investigate the impact of LCTs on LV feeders. Combined with the introduction of several visualization methods, it allows DSOs with access to smart meter data to gain insight in the qualitative and quantitative behavior of adding PV, EVs and HPs on their grid. We were able to use a unique large dataset of over 42,089 profiles, with many LCTs: 862 HPs, 1924 EVs, and 76\% PV. These were all measured in 2022 and represent actual consumption behavior of residential connections on the LVG in the region of Flanders, Belgium. This is the first time such large, recently measured dataset of LVG customers has been analyzed.  

Our feeder modeling approach makes a few simplifying assumptions. We assume that the profiles are a representative sample of all connections in Flanders, which is likely not true. Furthermore, the change in feeder peak when adding an LCT is estimated from the difference between feeders without any LCT and feeders where all connections have an LCT. Modeling without these assumptions is possible, but is significantly more complicated, as the properties of each connection should be taken into account in the model, see e.g.~\cite{SOENEN2023100985}. 

A large variation of behaviors was observed at the level of an individual connection. Because we had access to a large and diverse dataset, this variation in behaviors did not need to be modeled.
The distribution of the time the feeder peak occurs is wide, especially for small feeders of 10 connections.
The contribution of LCTs to yearly feeder peaks was significant: for a feeder with 40 connections a HP adds on average 1.2 kW, an EV adds on average 1.4 kW, and an EV charging faster than 6.5 kW adds 2.0 kW per connection. Injection peaks are of similar magnitude than offtake peaks. Whether the future (Flemish) LVG will be mostly offtake constrained of injection constrained (or a mix of the two) is an open question.

Feeders without HPs and without EVs exhibited classical duck-curve behavior. Feeders with HPs showed a night-camel curve during winter, with a morning- and evening peak of similar magnitude and a high camel back at night. The timing of the peak was highly temperature driven, and mostly occurs in the morning. For EVs that charge at high power ($>$ 6.5 kW), we observe a night-dromedary curve, with a large peak (hump) between 22h and 1h and a smaller morning peak (head).

Options for future research include analyzing the behavior of batteries, for which we didn't have labels, estimating the effect of the capacity tariff that was introduced in 2023, and analyzing connections that have both an EV and a HP (of which there weren't enough profiles in the current dataset). 

The results in this paper represent a snapshot of early adaptors of LCT technology in 2022 in Flanders. The behavior of customers with LCT devices is expected to evolve as a function of several factors such as new tariff structures and policy measures. Our introduced method can be used to follow-up on changes in consumption behavior.

\section*{CRediT authorship contribution statement}

\textbf{Thijs Becker}: Conceptualization, Methodology, Software, Investigation, Writing - Original Draft, Writing - Review \& Editing, Visualisation \textbf{Raf Smet}: Conceptualization, Data Curation, Writing - Review \& Editing \textbf{Bruno Macharis}: Conceptualization, Writing - Review \& Editing, Supervision, Project administration \textbf{Koen Vanthournout}: Conceptualization, Writing - Review \& Editing, Supervision, Project administration, Funding acquisition

\section*{Declaration of Competing Interest}

The authors declare that they have no known competing financial interest.

\section*{Data Availability}

Because of privacy reasons, the data can not be made available.

\section*{Acknowledgments}

We thank Fluvius for providing the data and supporting this publication. We thank Jan Yperman for providing the Flemish-level metadata.

\section*{Funding sources}

This research was performed during the ADriaN project, which is funded by Fluvius.

\section*{Declaration of Generative AI and AI-assisted technologies in the writing process}

During the preparation of this work the authors used Dall-E 3 in order to generate the night-dromedary included in the graphical abstract. After using this tool, the authors reviewed and edited the content as needed and take full responsibility for the content of the publication.

\appendix

\section{Day of year plots\label{app:doy}}


\begin{figure}[ht]
    \centering
    \begin{subfigure}[b]{0.495\textwidth}
        \includegraphics[width=\textwidth]{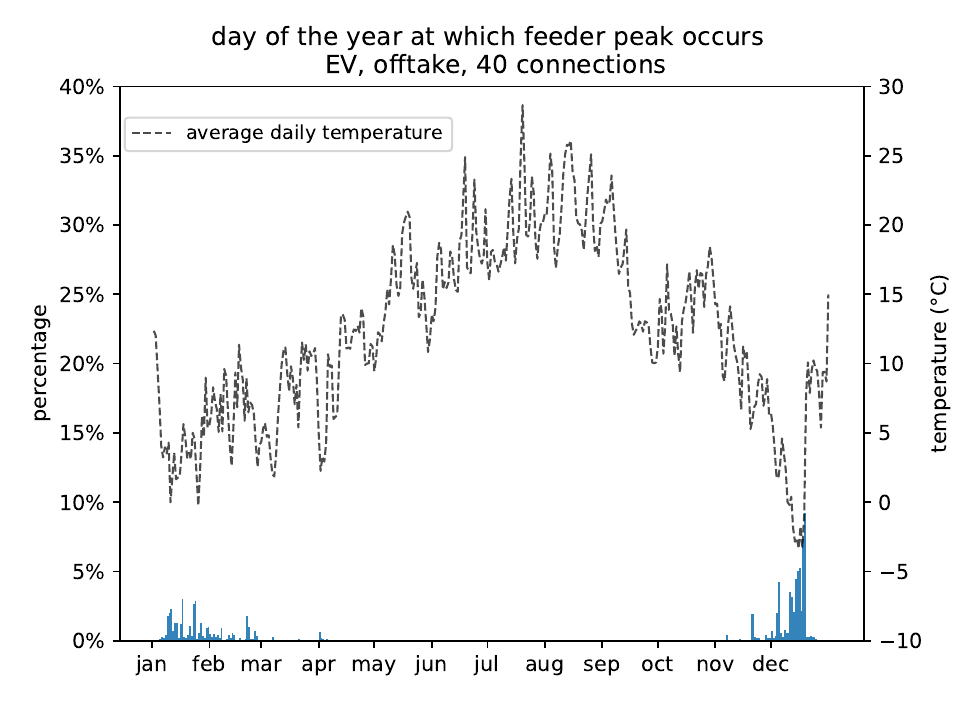}
        \caption{EV}
        \label{fig:ev_doy}
    \end{subfigure}
    \begin{subfigure}[b]{0.495\textwidth}
        \includegraphics[width=\textwidth]{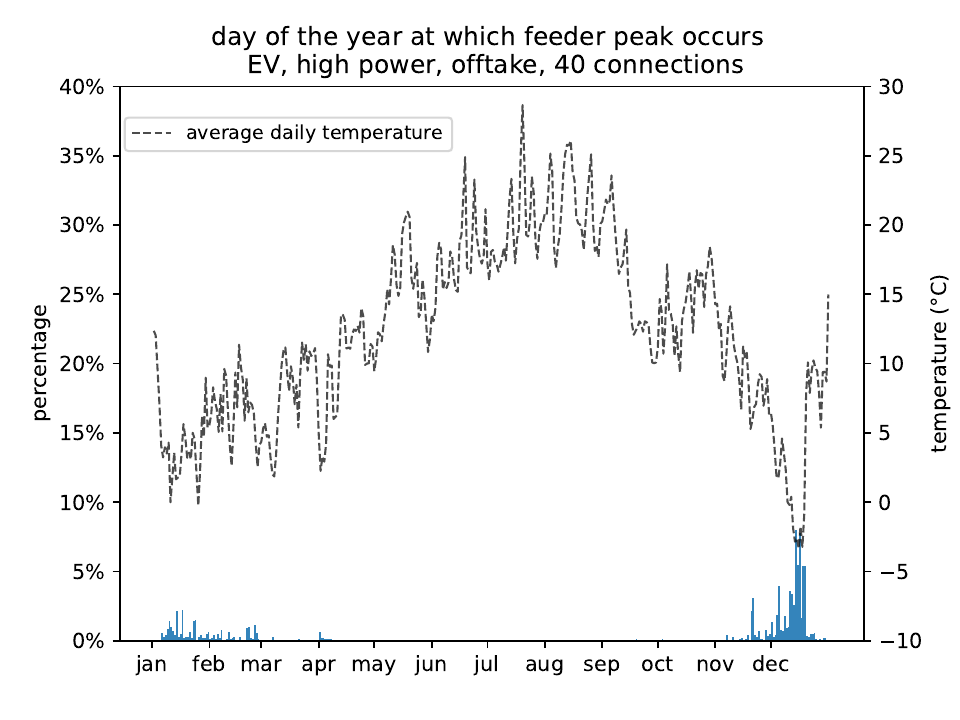}
        \caption{EV high}
        \label{fig:evhigh_doy}
    \end{subfigure}

    \caption{Histogram of on what day of the year the feeder peak occurs, for EV and EV high power, for 40 connections, with the average daily temperature.}
    \label{fig:doy_ev_evhigh}
\end{figure}

\begin{figure}[ht]
    \centering
    \begin{subfigure}[b]{0.495\textwidth}
        \includegraphics[width=\textwidth]{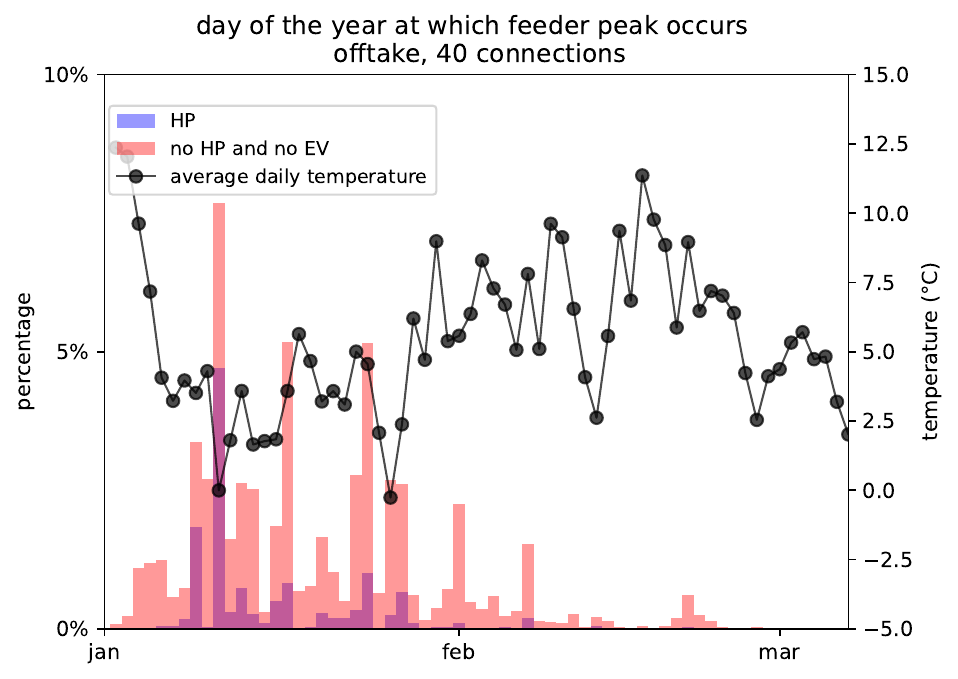}
        \caption{Start of the year.}
        \label{fig:zoomleft}
    \end{subfigure}
    \begin{subfigure}[b]{0.495\textwidth}
        \includegraphics[width=\textwidth]{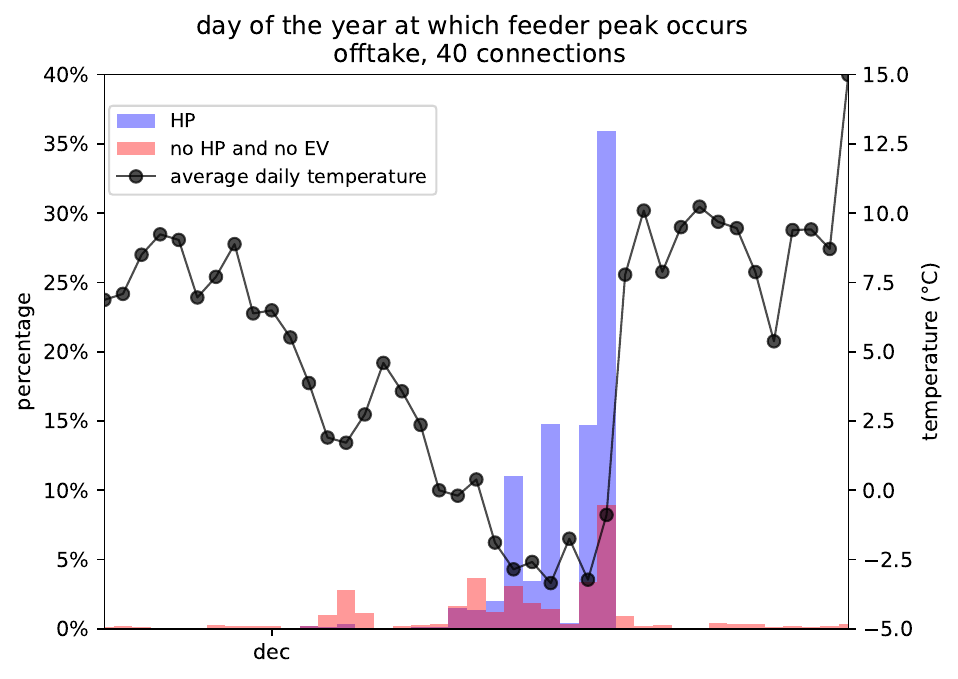}
        \caption{End of the year.}
        \label{fig:zoomright}
    \end{subfigure}

    \caption{Histogram of on what day of the year the feeder peak occurs, for HP and no HP and EV, for 40 connections. We have zoomed in on the months in which the feeder peaks occur}
    \label{fig:doy_wp_nowp_zoom}
\end{figure}


 \bibliographystyle{elsarticle-num} 
 \bibliography{references}

\begin{thebibliography}{10}
\expandafter\ifx\csname url\endcsname\relax
  \def\url#1{\texttt{#1}}\fi
\expandafter\ifx\csname urlprefix\endcsname\relax\def\urlprefix{URL }\fi
\expandafter\ifx\csname href\endcsname\relax
  \def\href#1#2{#2} \def\path#1{#1}\fi

\bibitem{IPCC2023}
{Core Writing Team}, H.~Lee, J.~Romero (Eds.), Summary for Policymakers. In:
  Climate Change 2023: Synthesis Report. Contribution of Working Groups I, II
  and III to the Sixth Assessment Report of the Intergovernmental Panel on
  Climate Change, IPCC, Geneva, Switzerland, 2023.
\newblock \href {https://doi.org/10.59327/IPCC/AR6-9789291691647.00}
  {\path{doi:10.59327/IPCC/AR6-9789291691647.00}}.

\bibitem{VOHRA2021110754}
K.~Vohra, A.~Vodonos, J.~Schwartz, E.~A. Marais, M.~P. Sulprizio, L.~J.
  Mickley,
  \href{https://www.sciencedirect.com/science/article/pii/S0013935121000487}{Global
  mortality from outdoor fine particle pollution generated by fossil fuel
  combustion: Results from geos-chem}, Environmental Research 195 (2021)
  110754.
\newblock \href {https://doi.org/https://doi.org/10.1016/j.envres.2021.110754}
  {\path{doi:https://doi.org/10.1016/j.envres.2021.110754}}.
\newline\urlprefix\url{https://www.sciencedirect.com/science/article/pii/S0013935121000487}

\bibitem{iea2023outlook}
IEA, \href{https://www.iea.org/reports/world-energy-outlook-2023}{World energy
  outlook}, IEA, Paris, licence: CC BY 4.0 (2023).
\newline\urlprefix\url{https://www.iea.org/reports/world-energy-outlook-2023}

\bibitem{eurostat2020}
Eurostat,
  \href{https://www.weforum.org/agenda/2022/09/eu-greenhouse-gas-emissions-transport/}{Greenhouse
  gas emissions by source sector, eu, 2020} (2022).
\newline\urlprefix\url{https://www.weforum.org/agenda/2022/09/eu-greenhouse-gas-emissions-transport/}

\bibitem{DAMIANAKIS2023}
N.~Damianakis, G.~R.~C. Mouli, P.~Bauer, Y.~Yu,
  \href{https://www.sciencedirect.com/science/article/pii/S0306261923012424}{Assessing
  the grid impact of electric vehicles, heat pumps \& pv generation in dutch lv
  distribution grids}, Applied Energy 352 (2023) 121878.
\newblock \href
  {https://doi.org/https://doi.org/10.1016/j.apenergy.2023.121878}
  {\path{doi:https://doi.org/10.1016/j.apenergy.2023.121878}}.
\newline\urlprefix\url{https://www.sciencedirect.com/science/article/pii/S0306261923012424}

\bibitem{li2024impact}
Y.~Li, A.~Jenn, Impact of electric vehicle charging demand on power
  distribution grid congestion, Proceedings of the National Academy of Sciences
  121~(18) (2024) e2317599121.

\bibitem{GAUNT20171}
C.~Gaunt, E.~Namanya, R.~Herman,
  \href{https://www.sciencedirect.com/science/article/pii/S0378779616303467}{Voltage
  modelling of lv feeders with dispersed generation: Limits of penetration of
  randomly connected photovoltaic generation}, Electric Power Systems Research
  143 (2017) 1--6.
\newblock \href {https://doi.org/https://doi.org/10.1016/j.epsr.2016.08.042}
  {\path{doi:https://doi.org/10.1016/j.epsr.2016.08.042}}.
\newline\urlprefix\url{https://www.sciencedirect.com/science/article/pii/S0378779616303467}

\bibitem{Elmallah_2022}
S.~Elmallah, A.~M. Brockway, D.~Callaway, Can distribution grid infrastructure
  accommodate residential electrification and electric vehicle adoption in
  northern california?, Environmental Research: Infrastructure and
  Sustainability 2~(4) (2022) 045005.
\newblock \href {https://doi.org/10.1088/2634-4505/ac949c}
  {\path{doi:10.1088/2634-4505/ac949c}}.

\bibitem{en13195083}
B.~Thormann, T.~Kienberger,
  \href{https://www.mdpi.com/1996-1073/13/19/5083}{Evaluation of grid
  capacities for integrating future e-mobility and heat pumps into low-voltage
  grids}, Energies 13~(19) (2020).
\newblock \href {https://doi.org/10.3390/en13195083}
  {\path{doi:10.3390/en13195083}}.
\newline\urlprefix\url{https://www.mdpi.com/1996-1073/13/19/5083}

\bibitem{iea2023Grids}
IEA,
  \href{https://www.iea.org/reports/electricity-grids-and-secure-energy-transitions}{Electricity
  grids and secure energy transitions}, IEA, Paris, licence: CC BY 4.0 (2023).
\newline\urlprefix\url{https://www.iea.org/reports/electricity-grids-and-secure-energy-transitions}

\bibitem{HABEN2021117798}
S.~Haben, S.~Arora, G.~Giasemidis, M.~Voss, D.~{Vukadinović Greetham},
  \href{https://www.sciencedirect.com/science/article/pii/S0306261921011326}{Review
  of low voltage load forecasting: Methods, applications, and recommendations},
  Applied Energy 304 (2021) 117798.
\newblock \href
  {https://doi.org/https://doi.org/10.1016/j.apenergy.2021.117798}
  {\path{doi:https://doi.org/10.1016/j.apenergy.2021.117798}}.
\newline\urlprefix\url{https://www.sciencedirect.com/science/article/pii/S0306261921011326}

\bibitem{VELDMAN2013}
E.~Veldman, M.~Gibescu, H.~J. Slootweg, W.~L. Kling,
  \href{https://www.sciencedirect.com/science/article/pii/S0301421513000189}{Scenario-based
  modelling of future residential electricity demands and assessing their
  impact on distribution grids}, Energy Policy 56 (2013) 233--247.
\newblock \href {https://doi.org/https://doi.org/10.1016/j.enpol.2012.12.078}
  {\path{doi:https://doi.org/10.1016/j.enpol.2012.12.078}}.
\newline\urlprefix\url{https://www.sciencedirect.com/science/article/pii/S0301421513000189}

\bibitem{era5}
J.~Mu\~noz Sabater, E.~Dutra, A.~Agust\'{\i}-Panareda, C.~Albergel, G.~Arduini,
  G.~Balsamo, S.~Boussetta, M.~Choulga, S.~Harrigan, H.~Hersbach, B.~Martens,
  D.~G. Miralles, M.~Piles, N.~J. Rodr\'{\i}guez-Fern\'andez, E.~Zsoter,
  C.~Buontempo, J.-N. Th\'epaut,
  \href{https://essd.copernicus.org/articles/13/4349/2021/}{Era5-land: a
  state-of-the-art global reanalysis dataset for land applications}, Earth
  System Science Data 13~(9) (2021) 4349--4383.
\newblock \href {https://doi.org/10.5194/essd-13-4349-2021}
  {\path{doi:10.5194/essd-13-4349-2021}}.
\newline\urlprefix\url{https://essd.copernicus.org/articles/13/4349/2021/}

\bibitem{LOVE2017}
J.~Love, A.~Z. Smith, S.~Watson, E.~Oikonomou, A.~Summerfield, C.~Gleeson,
  P.~Biddulph, L.~F. Chiu, J.~Wingfield, C.~Martin, A.~Stone, R.~Lowe,
  \href{https://www.sciencedirect.com/science/article/pii/S0306261917308954}{The
  addition of heat pump electricity load profiles to gb electricity demand:
  Evidence from a heat pump field trial}, Applied Energy 204 (2017) 332--342.
\newblock \href
  {https://doi.org/https://doi.org/10.1016/j.apenergy.2017.07.026}
  {\path{doi:https://doi.org/10.1016/j.apenergy.2017.07.026}}.
\newline\urlprefix\url{https://www.sciencedirect.com/science/article/pii/S0306261917308954}

\bibitem{barteczko2015after}
C.~Barteczko-Hibbert, After diversity maximum demand (admd) report, Report for
  the ‘Customer-Led Network Revolution’project: Durham University (2015).

\bibitem{veldman2011}
E.~Veldman, M.~Gibescu, H.~Slootweg, W.~L. Kling, Impact of electrification of
  residential heating on loading of distribution networks, in: 2011 IEEE
  Trondheim PowerTech, 2011, pp. 1--7.
\newblock \href {https://doi.org/10.1109/PTC.2011.6019179}
  {\path{doi:10.1109/PTC.2011.6019179}}.

\bibitem{Bollerslev2022}
J.~Bollerslev, P.~B. Andersen, T.~V. Jensen, M.~Marinelli, A.~Thingvad,
  L.~Calearo, T.~Weckesser, Coincidence factors for domestic ev charging from
  driving and plug-in behavior, IEEE Transactions on Transportation
  Electrification 8~(1) (2022) 808--819.
\newblock \href {https://doi.org/10.1109/TTE.2021.3088275}
  {\path{doi:10.1109/TTE.2021.3088275}}.

\bibitem{GONZALEZVENEGAS2021}
F.~{Gonzalez Venegas}, M.~Petit, Y.~Perez,
  \href{https://www.sciencedirect.com/science/article/pii/S2590116821000291}{Plug-in
  behavior of electric vehicles users: Insights from a large-scale trial and
  impacts for grid integration studies}, eTransportation 10 (2021) 100131.
\newblock \href {https://doi.org/https://doi.org/10.1016/j.etran.2021.100131}
  {\path{doi:https://doi.org/10.1016/j.etran.2021.100131}}.
\newline\urlprefix\url{https://www.sciencedirect.com/science/article/pii/S2590116821000291}

\bibitem{RICHARDSON20101878}
I.~Richardson, M.~Thomson, D.~Infield, C.~Clifford,
  \href{https://www.sciencedirect.com/science/article/pii/S0378778810001854}{Domestic
  electricity use: A high-resolution energy demand model}, Energy and Buildings
  42~(10) (2010) 1878--1887.
\newblock \href {https://doi.org/https://doi.org/10.1016/j.enbuild.2010.05.023}
  {\path{doi:https://doi.org/10.1016/j.enbuild.2010.05.023}}.
\newline\urlprefix\url{https://www.sciencedirect.com/science/article/pii/S0378778810001854}

\bibitem{flett2022_modelica_admd}
G.~Flett, P.~Tuohy, Towards a high-resolution, stochastic, domestic energy
  demand model to assess the local network impact of heat and transport
  electrification, in: uSIM 2022 Conference: Urban Energy in a Net Zero World,
  2022, pp. 1--9.

\bibitem{PROTOPAPADAKI2017268}
C.~Protopapadaki, D.~Saelens,
  \href{https://www.sciencedirect.com/science/article/pii/S0306261916317329}{Heat
  pump and pv impact on residential low-voltage distribution grids as a
  function of building and district properties}, Applied Energy 192 (2017)
  268--281.
\newblock \href
  {https://doi.org/https://doi.org/10.1016/j.apenergy.2016.11.103}
  {\path{doi:https://doi.org/10.1016/j.apenergy.2016.11.103}}.
\newline\urlprefix\url{https://www.sciencedirect.com/science/article/pii/S0306261916317329}

\bibitem{pflugradt2013analysing}
N.~Pflugradt, J.~Teuscher, B.~Platzer, W.~Schufft, Analysing low-voltage grids
  using a behaviour based load profile generator, in: International conference
  on renewable energies and power quality, Vol.~11, 2013, p.~5.

\bibitem{synthLiang2022}
X.~Liang, H.~Wang,
  \href{https://www.climatechange.ai/events/neurips2022}{Synthesis of realistic
  load data: adversarial networks for learning and generating residential load
  patterns}, in: P.~Mitra, M.~{Jo{\~a}o Sousa}, M.~Roth, J.~Drgo{\v n}a,
  E.~Strubell, Y.~Bengio (Eds.), NeurIPS 2022 Workshop, Neural Information
  Processing Systems (NIPS), 2022, pp. 1--8, tackling Climate Change with
  Machine Learning 2022 ; Conference date: 09-12-2022 Through 09-12-2022.
\newline\urlprefix\url{https://www.climatechange.ai/events/neurips2022}

\bibitem{synthGu2019}
Y.~Gu, Q.~Chen, K.~Liu, L.~Xie, C.~Kang, Gan-based model for residential load
  generation considering typical consumption patterns, in: 2019 IEEE Power \&
  Energy Society Innovative Smart Grid Technologies Conference (ISGT), 2019,
  pp. 1--5.
\newblock \href {https://doi.org/10.1109/ISGT.2019.8791575}
  {\path{doi:10.1109/ISGT.2019.8791575}}.

\bibitem{energiewandlung2008vdi}
V.~Energiewandlung, Vdi 4655—reference load profiles of single-family and
  multi-family houses for the use of chp systems, Tech. Guidel 11 (2008) 123.

\bibitem{ukdataset}
U.~P. networks, {Smartmeter energy consumption data in london houseolds},
  \url{https://data.london.gov.uk/dataset/
  smartmeter-energy-use-data-in-london-households} (2018).

\bibitem{WANG2020116780}
Z.~Wang, J.~Crawley, F.~G. Li, R.~Lowe,
  \href{https://www.sciencedirect.com/science/article/pii/S0360544219324752}{Sizing
  of district heating systems based on smart meter data: Quantifying the
  aggregated domestic energy demand and demand diversity in the uk}, Energy 193
  (2020) 116780.
\newblock \href {https://doi.org/https://doi.org/10.1016/j.energy.2019.116780}
  {\path{doi:https://doi.org/10.1016/j.energy.2019.116780}}.
\newline\urlprefix\url{https://www.sciencedirect.com/science/article/pii/S0360544219324752}

\bibitem{irish_ashp_2021}
P.~L. Michael~Chesser, Padraic~O'Reilly, P.~Carroll,
  \href{https://doi.org/10.1080/15567249.2021.1945169}{The impact of extreme
  weather on peak electricity demand from homes heated by air source heat
  pumps}, Energy Sources, Part B: Economics, Planning, and Policy 16~(8) (2021)
  707--718.
\newblock \href
  {http://arxiv.org/abs/https://doi.org/10.1080/15567249.2021.1945169}
  {\path{arXiv:https://doi.org/10.1080/15567249.2021.1945169}}, \href
  {https://doi.org/10.1080/15567249.2021.1945169}
  {\path{doi:10.1080/15567249.2021.1945169}}.
\newline\urlprefix\url{https://doi.org/10.1080/15567249.2021.1945169}

\bibitem{SILBER2024}
F.~Silber, S.~Scheubner, A.~Märtz,
  \href{https://www.sciencedirect.com/science/article/pii/S0142061523005975}{Analysis
  of the simultaneity factor of fast-charging sites using monte-carlo
  simulation}, International Journal of Electrical Power \& Energy Systems 155
  (2024) 109540.
\newblock \href {https://doi.org/https://doi.org/10.1016/j.ijepes.2023.109540}
  {\path{doi:https://doi.org/10.1016/j.ijepes.2023.109540}}.
\newline\urlprefix\url{https://www.sciencedirect.com/science/article/pii/S0142061523005975}

\bibitem{Fani2023}
H.~Fani, M.~U. Hashmi, G.~Deconinck, Impact of electric vehicle charging
  simultaneity factor on the hosting capacity of lv feeder, SSRN, available at
  SSRN: \url{https://ssrn.com/abstract=4865750} or
  \url{http://dx.doi.org/10.2139/ssrn.4865750} (2024).

\bibitem{hungbo2023impact}
M.~Hungbo, M.~Gu, L.~Meegahapola, T.~Littler, S.~Bu, Impact of electric
  vehicles on low-voltage residential distribution networks: A probabilistic
  analysis, IET Smart Grid 6~(5) (2023) 536--548.

\bibitem{9667519}
Y.~Yu, D.~Reihs, S.~Wagh, A.~Shekhar, D.~Stahleder, G.~R.~C. Mouli, F.~Lehfuss,
  P.~Bauer, Data-driven study of low voltage distribution grid behaviour with
  increasing electric vehicle penetration, IEEE Access 10 (2022) 6053--6070.
\newblock \href {https://doi.org/10.1109/ACCESS.2021.3140162}
  {\path{doi:10.1109/ACCESS.2021.3140162}}.

\bibitem{admd_probabilistic_2019}
M.~Sun, Y.~Wang, G.~Strbac, C.~Kang, Probabilistic peak load estimation in
  smart cities using smart meter data, IEEE Transactions on Industrial
  Electronics 66~(2) (2019) 1608--1618.
\newblock \href {https://doi.org/10.1109/TIE.2018.2803732}
  {\path{doi:10.1109/TIE.2018.2803732}}.

\bibitem{admd_probablistic_2004}
D.~McQueen, P.~Hyland, S.~Watson, Monte carlo simulation of residential
  electricity demand for forecasting maximum demand on distribution networks,
  IEEE Transactions on Power Systems 19~(3) (2004) 1685--1689.
\newblock \href {https://doi.org/10.1109/TPWRS.2004.826800}
  {\path{doi:10.1109/TPWRS.2004.826800}}.

\bibitem{python_book}
G.~Van~Rossum, F.~L. Drake, Python 3 Reference Manual, CreateSpace, Scotts
  Valley, CA, 2009.

\bibitem{harris2020array}
C.~R. Harris, K.~J. Millman, S.~J. van~der Walt, R.~Gommers, P.~Virtanen,
  D.~Cournapeau, E.~Wieser, J.~Taylor, S.~Berg, N.~J. Smith, R.~Kern, M.~Picus,
  S.~Hoyer, M.~H. van Kerkwijk, M.~Brett, A.~Haldane, J.~F. del R{\'{i}}o,
  M.~Wiebe, P.~Peterson, P.~G{\'{e}}rard-Marchant, K.~Sheppard, T.~Reddy,
  W.~Weckesser, H.~Abbasi, C.~Gohlke, T.~E. Oliphant,
  \href{https://doi.org/10.1038/s41586-020-2649-2}{Array programming with
  {NumPy}}, Nature 585~(7825) (2020) 357--362.
\newblock \href {https://doi.org/10.1038/s41586-020-2649-2}
  {\path{doi:10.1038/s41586-020-2649-2}}.
\newline\urlprefix\url{https://doi.org/10.1038/s41586-020-2649-2}

\bibitem{BRUDERMUELLER2023}
T.~Brudermueller, M.~Kreft, E.~Fleisch, T.~Staake,
  \href{https://www.sciencedirect.com/science/article/pii/S030626192301098X}{Large-scale
  monitoring of residential heat pump cycling using smart meter data}, Applied
  Energy 350 (2023) 121734.
\newblock \href
  {https://doi.org/https://doi.org/10.1016/j.apenergy.2023.121734}
  {\path{doi:https://doi.org/10.1016/j.apenergy.2023.121734}}.
\newline\urlprefix\url{https://www.sciencedirect.com/science/article/pii/S030626192301098X}

\bibitem{denholm2015overgeneration}
P.~Denholm, M.~O'Connell, G.~Brinkman, J.~Jorgenson, Overgeneration from solar
  energy in california: A field guide to the duck chart, Tech. rep., National
  Renewable Energy Lab.(NREL), Golden, CO (United States) (2015).

\bibitem{SOENEN2023100985}
J.~Soenen, A.~Yurtman, T.~Becker, R.~D’hulst, K.~Vanthournout, W.~Meert,
  H.~Blockeel,
  \href{https://www.sciencedirect.com/science/article/pii/S2352467722002302}{Scenario
  generation of residential electricity consumption through sampling of
  historical data}, Sustainable Energy, Grids and Networks 34 (2023) 100985.
\newblock \href {https://doi.org/https://doi.org/10.1016/j.segan.2022.100985}
  {\path{doi:https://doi.org/10.1016/j.segan.2022.100985}}.
\newline\urlprefix\url{https://www.sciencedirect.com/science/article/pii/S2352467722002302}

\end{thebibliography}

\end{document}